\documentclass[preprint,aps,12pt,showpacs,nofootinbib,tightenlines]{revtex4}
\usepackage{mathrsfs}
\usepackage{amsmath}
\usepackage{amssymb}
\usepackage{epsfig}
\usepackage{epstopdf}
\usepackage{graphicx}
\usepackage{subfigure}
\usepackage{booktabs}
\usepackage{float}
\usepackage{amsmath}
\usepackage{color}
\usepackage{multirow}
\usepackage{bm}

\def\be{\begin{eqnarray}}
\def\en{\end{eqnarray}}
\def\non{\nonumber\\}

\begin{document}
%%--------------------------------------------
\title{ Semi-leptonic decays $B \to D^{(*)}(1S,2S)\ell \nu_{\ell}$ within the covariant light-front approach  }
\author{Zhi-Jie Sun$^1$, 
	Zhi-Qing Zhang$^2$\footnote{ Corresponding author. zhangzhiqing@haut.edu.cn}, Shi-Chen Xue$^2$, Feng-Zhou Wang$^1$} %%
\affiliation{\it \small $^1$ Bingtuan Xingxin Vocational and Technical College, Tiemenguan, Xinjiang, 841007, China\\
	\it \small $^2$ School of Physics and Advanced Energy, Henan University of Technology, Zhengzhou, Henan 450001, China } %%
\date{\today}
\begin{abstract}
We present a systematic analysis of the semi-leptonic decays $B_{(s)}\to D_{(s)}(1S,2S)\ell\nu_\ell$ and $B_{(s)}\to D^*_{(s)}(1S,2S)\ell\nu_\ell$ with $\ell=e,\mu,\tau$ within the covariant light-front quark model (CLFQM). Using the form factors of the transitions $B_{(s)}\to D_{(s)}(1S,2S)$ and $B_{(s)}\to D^*_{(s)}(1S,2S)$, we calculate the branching ratios of the relevant semi-leptonic decays and find that $Br(B_{(s)}\to D_{(s)}\ell^\prime\nu_{\ell^\prime})$ and $Br(B_{(s)}\to D^*_{(s)}\ell^\prime\nu_{\ell^\prime})$ with $\ell^\prime=e,\mu$ are agree well with the data, while $Br(B_{(s)}\to D_{(s)}\tau\nu_{\tau})$ and $Br(B_{(s)}\to D^*_{(s)}\tau\nu_{\tau})$ are systematically smaller than the experimental measurements. This naturally gives rise to the so-called $\mathcal{R}(D)$ and $\mathcal{R}(D^*)$ anomalies. Our predictions $\mathcal{R}(D)=0.261\pm0.013$ and $\mathcal{R}(D^*)=0.228\pm0.026$ show $3.1\sigma$ and $2.1\sigma$ deviations from the current experimental world averages compiled by the Heavy Flavor Averaging Group (HFLAV), respectively, yet only deviate by $0.16\sigma$ and $1.5\sigma$ from the latest LHCb measurements. For the decays 
$B_{(s)}\to D_{(s)}(2S)\ell\nu_\ell$ and $B_{(s)}\to D^*_{(s)}(2S)\ell\nu_\ell$, their branching ratios lie in the range $10^{-4}\sim10^{-3}$, which are much larger than the results from the Bethe Salpeter (BS) equation , but agree with the relativistic quark model (RQM) calculations. Furthermore, we also calculate the forward-backward asymmetries $\mathcal{A}_{FB}$ and longitudinal polarization fractions
$f_L$ for the corresponding decays. Our predictions are consistent with most other theoretical
results and experimental data.
\end{abstract}

 \vspace{1cm}

\maketitle

%=======================================================================
%                     Introduction
%=======================================================================
\section{Introduction}\label{intro}
The semi-leptonic decays of heavy flavor mesons play an central role in constraining the Standard Model (SM) and searching for the signal of New Physics (NP).
These decays provide an ideal platform for precise extraction of the Cabibbo-Kobayashi-Maskawa (CKM) matrix elements, the fundamental free parameters of the SM. The SM assumes that the electroweak couplings to all three generations of leptons are universal and the only difference arises due to the their mass differences, a principle known as lepton flavor universality (LFU). Thus, the lepton flavor symmetry between semileptonic decay rates can be obtained after considering the charged lepton mass contributions to decay amplitudes and phase space. Any deviations from the CKM matrix unitarity constraints or the LFU will be a signal of NP.

Over the past several years, the $b \to c\ell\nu_\ell$ transition has
served as a powerful precision probe for the search of NP, especially experimental measurements of the semi-leptonic process $b\to c\tau\nu_\tau$ yield branching ratios systematiclly larger than SM expectations, which could be a hint of LFU violation \cite{Bernlochner}.
The branching ratios of the decays $B\to D^{(*)}\ell\nu_\ell$ have been determined by the BaBar, Belle, and LHCb collaborations. To minimize hadronic uncertainties, these experiments further extract the ratios of the branching fractions of the semileptonic decays $B\to D^{(*)}\ell\nu_\ell$ \cite{lhcb1}, which will be discussed in detail in the numerical results section.
%\begin{eqnarray*}
%	\frac{\mathcal{B}r(B_s^0 \to D_s^- \mu^+ \nu_\mu)}{\mathcal{B}r(B^0 \to D^- \mu^+ \nu_\mu)} = 1.09 \pm 0.05_{\text{stat}} \pm 0.06_{\text{syst}} \pm 0.05_{\text{ext}},\\
%	\frac{\mathcal{B}r(B_s^0 \to D_s^{*-} \mu^+ \nu_\mu)}{\mathcal{B}r(B^0 \to D^{*-} \mu^+ \nu_\mu)} = 1.06 \pm 0.05_{\text{stat}} \pm 0.07_{\text{syst}} \pm 0.05_{\text{ext}}.
%\end{eqnarray*}

Of particular interest are the LFU ratios,
\be
\mathcal{R}\left(D\right)&=&\frac{B r\left(B \rightarrow D \tau \nu_\tau\right)}{Br\left(B \rightarrow D\ell^{\prime} \nu_{\ell^\prime}\right)},\quad
\mathcal{R}\left(D^{\ast}\right)=\frac{B r\left(B \rightarrow D^{\ast} \tau \nu_\tau\right)}{Br\left(B \rightarrow D^{\ast}\ell^{\prime} \nu_{\ell^\prime}\right)},
\en
which have been measured by the different collaborations and the latest world averages compiled by HFLAV are listed as
\be 
\mathcal{R}(D)_{exp}=0.347\pm0.025, \quad \mathcal{R}(D^*)_{exp}=0.288\pm0.012.
\en
where the results include both statistical and systematic uncertainties.
On the theoretical side, the transition form factors, branching fractions and related ratios are calculated in different approaches, such as
the three point QCD sum rules (QCDSR) \cite{Azizi:2008tt,Azizi:2008vt}, the light cone QCD sum rules (LCSR) \cite{Bordone:2019guc,Zhang:2022opp}, the relativistic quark model (RQM) \cite{Faustov:2022ybm} and so on.  To achieve more precise determinations of the $B_{(s)}\to D_{(s)}$ and $B_{(s)}\to D^*_{(s)}$ transition form factors, the authors of Ref. \cite{Cui:2023jiw} have calculated the next-to-leading order QCD corrections at large hadronic recoil.
Certainly, high-precision lattice QCD calculations for these transitions are also available in Refs. \cite{FermilabLattice:2015ilb,Na:2015kha2,Na:2015kha,Aviles-Casco:2019zop,Kaneko:2019vkx}. HFLAV provides the SM central values for the LFU ratios as
\be
\mathcal{R}(D)_{SM}=0.296\pm0.004, \quad \mathcal{R}(D^*)_{SM}=0.254\pm0.005,
\en
which are obtained from global fits that combining lattice QCD inputs at high-$q^2$ with
experimental form factor determinations at low-$q^2$ \cite{2024oxs,2023fwm,Ray:2023xjn}. These results rely on parameterization schemes to extrapolate form factors from a 
restricted $q^2$ region to the full kinematic $q^2$ range. One can find 
that the measured $\mathcal{R}(D)$ and 
$\mathcal{R}(D^*)$ values exceed the SM predictions given above by $2.0\sigma$ and $2.7\sigma$, respectively. 
Considering the $\mathcal{R}(D)-\mathcal{R}(D^*)$ correlation,  the 
deviation between the data and the SM predictions remains at about $3.8\sigma$. Though this is smaller than the earlier $4\sigma$ tension, it is still one of the most persistent hints of LFU violation in the $B$
sector. Various heavy flavor anomalies associated with these transitions can be found in Refs.\cite{Li:2018lxi,Albrecht:2021tul,London:2021lfn} and references therein.

The remainder of this paper is structured as follows. The formalism of the CLFQM, the hadronic matrix elements, and the helicity amplitudes combined via form factors for the $B\to D^{(*)}(1S,2S)\ell\nu_\ell$ \footnote{From now on, $D_{(s)}$ and $D^*_{(s)}$ refer to $D_{(s)}(1S,2S)$ and $D^*_{(s)}(1S,2S)$, respectively.} decays are given in Section \ref{form1}. Using the  $B_{(s)}\to D_{(s)}, D^*_{(s)}$ form factor results derived in our previous work \cite{Sun:2025yiz}, we calculate the branching ratios, the forward-backward asymmetries and the polarization fractions for the corresponding semileptonic decays, which are presented in Section \ref{form2}. Section \ref{sum} provides a brief summary.
\section{THEORETICAL FRAMEWORK}\label{form1}
\subsection{Covariant light-front approach}
In the covariant quark model, the treatment of transition form factors is relatively covariant throughout the calculation process, where the light-front coordinates of a momentum $p$ are used $p=(p^-,p^+,p_\perp)$ with
$p^\pm=p^0\pm p_z, p^2=p^+p^--p^2_\perp$.
The incoming (outgoing) meson has the mass $M^\prime(M^{\prime\prime})$
with the momentum $P^\prime=p_1^\prime+p_2 (P^{\prime\prime}=p_1^{\prime\prime}+p_2)$, where
$p_{1}^{\prime(\prime\prime)}$
and $p_{2}$ are the momenta of the quark and anti-quark
inside the incoming (outgoing) meson with the mass $m_{1}^{\prime(\prime\prime)}$and $m_{2}$, respectively. Here we follow the notation covention
established in Refs. \cite{jaus,Y. Cheng}.
These momenta can be expressed in terms of the internal variables $(x_{i},p{'}_{\perp})$ as
\be
p_{1,2}^{\prime+}=x_{1,2} P^{\prime+}, \quad p_{1,2 \perp}^{\prime}=x_{1,2} P_{\perp}^{\prime} \pm p_{\perp}^{\prime}
\en
with $x_{1}+x_{2}=1$. Using these internal variables,
we can define some quantities for the incoming meson which relevant to subsequent calculations
\be
M_{0}^{\prime 2} &=&\left(e_{1}^{\prime}+e_{2}\right)^{2}=\frac{p_{\perp}^{\prime 2}+m_{1}^{\prime 2}}{x_{1}}
+\frac{p_{\perp}^{2}+m_{2}^{2}}{x_{2}}, \quad \widetilde{M}_{0}^{\prime}=\sqrt{M_{0}^{\prime 2}-\left(m_{1}^{\prime}-m_{2}\right)^{2}},\non
e_{i}^{(\prime)} &=&\sqrt{m_{i}^{(\prime) 2}+p_{\perp}^{\prime 2}+p_{z}^{\prime 2}}, \quad \quad p_{z}^{\prime}
=\frac{x_{2} M_{0}^{\prime}}{2}-\frac{m_{2}^{2}+p_{\perp}^{\prime 2}}{2 x_{2} M_{0}^{\prime}},
\en
where $M^\prime_0$ is the kinetic invariant mass of the incoming meson and can be expressed as the energies of the quark and the anti-quark
$e^{(\prime)}_i$. Identical expressions hold for the outgoing meson, but with $\prime$ replaced by $\prime\prime$ in superscripts.
\subsection{Form factors}
The Bauer-Stech-Wirble (BSW) form factors for the $B \rightarrow D$ and  $B \rightarrow D^{*}$ transitions are defined as follows
\begin{footnotesize}
	\begin{eqnarray}
		\left\langle D\left(P^{\prime
			\prime}\right)\left|V_{\mu}\right|
		B\left(P^{\prime}\right)\right\rangle
		&=&\left(P_{\mu}-\frac{m_{B}^{2}-m_{D}^{2}}{q^{2}}
		q_{\mu}\right) F_{1}\left(q^{2}\right)+\frac{m_B^2-m_{D}^2}{q^{2}} q_{\mu} F_{0}\left(q^{2}\right),\\    
		\label{pdep} 
		\left\langle D^{*}\left(P^{\prime \prime},\epsilon^{\prime\prime*}\right)\left|V_{\mu}-A_{\mu}\right| B \left(P^{\prime}\right)\right\rangle
		&=&-\epsilon_{\mu \nu \alpha \beta} \epsilon^{\prime\prime*}_\mu P^{\alpha} q^{\beta} \frac{V\left(q^{2}\right)}{m_{B}+m_{D^{*}}}-i \frac{2 m_{D^*} \epsilon^{\prime\prime*}\cdot P}{q^{2}} q_{\mu} A_{0}\left(q^{2}\right) \nonumber \\
		&&-i \epsilon^{\prime\prime*}_{\mu}\left(m_{B}+m_{D^{*}}\right) A_{1}\left(q^{2}\right)+i \frac{\epsilon^{\prime\prime*} \cdot P}{m_{B}+m_{D^{*}}}P_{\mu} A_{2}\left(q^{2}\right) \nonumber \\
		&&+i \frac{2 m_{D^{*}} \epsilon^{\prime\prime*} \cdot P}{q^{2}} q_{\mu} A_{3}\left(q^{2}\right),\label{pdev}
	\end{eqnarray}
\end{footnotesize}
where $P=P'+P'', q=P'-P''$ and the convention $\epsilon_{0123}=1$ is adopted. In order to calculate the amplitudes of the transition form factors, we need the following Feynman rules for the meson-quark-antiquark vertex $i\Gamma^{\prime(\prime\prime)} _{M}$ with $M$ representing a pseudoscalar (P) or vector (V) meson,
\be
i\Gamma^{\prime} _{P}&=&H^{\prime}_{P}\gamma_{5},\\
i\Gamma^{\prime\prime} _{P}&=&\gamma_0H^{\prime\prime}_{P}\gamma_{5}\gamma_0,\\
i\Gamma^{\prime\prime} _{V}&=&i \gamma_0H_{V}^{\prime\prime}\left[\gamma_{\mu}-\frac{1}{W_{V}^{\prime\prime}}\left(p_{1}^{\prime\prime}-p_{2}\right)_{\mu}\right]\gamma_0.
\en
\begin{figure}
	\subfigure{
		\begin{minipage}{6cm}
			\centering
			\includegraphics[width=6cm]{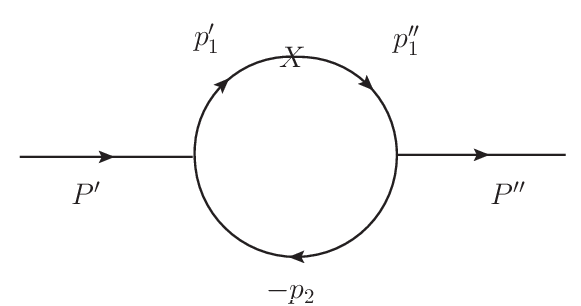}
	\end{minipage}}
	\caption{Feynman diagram for the meson weak transition, where X
		denotes the vector or axial-vector transition vertex.}
	\label{feyn}
\end{figure}
The form factors could be obtained by calculating the Feynman diagrams shown in Figure \ref{feyn}. 
For the general $P\to P$ transition, the amplitude for the lowest order is
\be
\mathcal{A}_{\mu}^{P P}=-i^{3} \frac{N_{c}}{(2 \pi)^{4}} \int d^{4} p_{1}^{\prime} \frac{H_{P}^{\prime}H_{P}^{\prime\prime}}
{N_{1}^{\prime} N_{1}^{\prime \prime} N_{2}} S_{\mu}^{P P},
\label{ZF}
\en
where $N_{1}^{\prime(\prime \prime)}=p_{1}^{\prime(\prime \prime) 2}-m_{1}^{\prime (\prime\prime) 2}$ and $ N_{2}=p_{2}^{2}-m_{2}^{2} $ arise
from the quark propagators, and
the trace $S_{\mu}^{PP}$ can be obtained directly by using the Lorentz contraction,
\be
S_{\mu}^{P P}&=&\operatorname{Tr}\left[\gamma_{5}\left(\not p_{1}^{\prime \prime}+m_{1}^{\prime \prime}\right) \gamma_{\mu}\left(\not p_{1}^{\prime}
+m_{1}^{\prime}\right) \gamma_{5}\left(-\not p_{2}+m_{2}\right)\right],
\label{ptop}
\en
where the analytical expression for $S_{\mu}^{P P}$ can be found in Ref. \cite{Zhang:2023ypl}.  It is similar for the $P\to V$ transition amplitude,
\be
\mathcal{A}_{\mu}^{P V}=-i^{3} \frac{N_{c}}{(2 \pi)^{4}} \int d^{4} p_{1}^{\prime} \frac{H_{P}^{\prime}\left(i H_{V}^{\prime \prime}\right)}{N_{1}^{\prime} N_{1}^{\prime \prime} N_{2}}
S_{\mu \nu}^{P V} \varepsilon^{*\nu},
\en
where
\be
S_{\mu \nu}^{P V}&=&\left(S_{V}^{P V}-S_{A}^{P V}\right)_{\mu \nu}\non
&=&\operatorname{Tr}\left[\left(\gamma_{\nu}-\frac{1}{W_{V}^{\prime \prime}}\left(p_{1}^{\prime \prime}-p_{2}\right)_{\nu}\right)\left(p_{1}^{\prime \prime}
+m_{1}^{\prime \prime}\right)\left(\gamma_{\mu}-\gamma_{\mu} \gamma_{5}\right)\left(\not p_{1}^{\prime}+m_{1}^{\prime}\right) \gamma_{5}\left(-\not p_{2}
+m_{2}\right)\right].\;\;\;
\label{sptov}
\en

In practice, we use the light-front decomposition of the Feynman loop momentum and integrate out
the minus component through the contour method. If the covariant vertex functions are not singular when performing integration,
the transition amplitudes will
pick up the singularities in the anti-quark propagators. The integration then leads to
\be
N_{1}^{\prime(\prime \prime)} &\rightarrow& \hat{N}_{1}^{\prime(\prime \prime)}=x_{1}\left(M^{\prime(\prime \prime) 2}-M_{0}^{\prime(\prime \prime) 2}\right),\non
\int \frac{d^{4} p_{1}^{\prime}}{N_{1}^{\prime} N_{1}^{\prime \prime} N_{2}} H_{P}^{\prime} H_{M}^{\prime \prime} S^{PM} & \rightarrow&-i \pi \int \frac{d x_{2} d^{2}
	p_{\perp}^{\prime}}{x_{2} \hat{N}_{1}^{\prime} \hat{N}_{1}^{\prime \prime}} h_{P}^{\prime} h_{M}^{\prime \prime} \hat{S}^{PM},\non
W_{V}^{\prime \prime} &\rightarrow& w_{V}^{\prime \prime}=M^{\prime\prime}_{0}+m^{\prime\prime}_{1}+m_{2},    
\en
where
\be
M_{0}^{\prime \prime 2}=\frac{p_{\perp}^{\prime \prime 2}+m_{1}^{\prime \prime 2}}{x_{1}}+\frac{p_{\perp}^{\prime \prime 2}+m_{2}^{2}}{x_{2}},
\label{vertex}
\en
with $p''_\perp=p'_\perp-x_2q_\perp$. The explicit forms of $h^{\prime\prime}_{M}$ are given as \cite{Y. Cheng}
\be
h_{P}^{\prime\prime} &=&h_{V}^{\prime\prime}=\left(M^{\prime\prime 2}-M_{0}^{\prime\prime 2}\right) \sqrt{\frac{x_{1} x_{2}}{N_{c}}} \frac{1}{\sqrt{2} \widetilde{M}_{0}^{\prime}} \varphi^{\prime\prime},
\label{hp}
\en
where $\varphi^{\prime\prime}$ is the light-front momentum distribution amplitude for S-wave mesons,
\be
\varphi^{\prime\prime} &=&\varphi^{\prime\prime}\left(x_{2}, p_{\perp}^{\prime\prime}\right)=4\left(\frac{\pi}{\beta^{2}}\right)^{\frac{3}{4}}
\sqrt{\frac{d p_{z}^{\prime\prime}}{d x_{2}}} \exp \left(-\frac{p_{z}^{\prime\prime 2}+p_{\perp}^{\prime\prime 2}}{2 \beta^{2}}\right).
\en
Here $\beta$ is a phenomenological parameter and can be fixed by fitting the corresponding decay constant. For the first radially excited charmed mesons $D_{(s)}^{*}(2S)$ and $D_{(s)}(2S)$, the distribution amplitudes are given as
\be
\varphi^{\prime\prime}(2S) &=&4\left(\frac{\pi}{\beta^{2}}\right)^{\frac{3}{4}}
\sqrt{\frac{d p_{z}^{\prime\prime}}{d x_{2}}} \exp \left(-\frac{p_{z}^{\prime\prime 2}+p_{\perp}^{\prime\prime 2}}{2 \beta^{2}}\right)
\times \frac{1}{\sqrt{6}}\left(-3+2\frac{p_{z}^{\prime\prime 2}+p_{\perp}^{\prime\prime 2}}{\beta^{2}}\right).
\en
%Using the formulas provided above and taking the integration rules given in Refs \cite{jaus,Y. Cheng},
%we obtain the expressions of the $B_{(s)}\to D_{(s)}(1S,2S)$ and $B_{(s)}\to D^*_{(s)}(1S,2S)$ transition form factors, which are listed in Appendix B.
\subsection{ Helicity amplitudes and observables}
The differential decay widths of the semileptonic decays $B_{(s)}\to D_{(s)}\ell\nu_\ell$ and $B_{(s)}\to D^*_{(s)}\ell\nu_\ell$  can be obtained by the combinations of helicity amplitudes, which are listed as follows:
\begin{footnotesize}
	\begin{align*}
		\frac{d\Gamma(B \to D \ell\nu)}{dq^2} &= (1 - \hat{m}_\ell^2)^2 \frac{\sqrt{\lambda(m_{B}^2, m_{D}^2, q^2)} G_F^2 |V_{cb}|^2}{384 m_{B}^3 \pi^3} \Bigg\{ \left(\hat{m}_\ell^2 + 2\right)\lambda(m_{B}^2, m_{D}^2, q^2) F_1^2(q^2) \label{ptop}\\
		& + 3\hat{m}_\ell^2 (m_{B}^2 - m_{D}^2)^2 F_0^2(q^2) \Bigg\}, \\[2pt]
		\frac{d\Gamma_L(B \to  D^*_{(s)} \ell\nu)}{dq^2} &= (1 - \hat{m}_\ell^2)^2 \frac{\sqrt{\lambda(m_{B}^2, m_{D^*}^2, q^2)} G_F^2 |V_{cb}|^2}{384 m_{B}^3 \pi^3} \Bigg\{ 3\hat{m}_\ell^2 \lambda(m_{B}^2, m_{D^*}^2, q^2) A_0^2(q^2)+ (\hat{m}_\ell^2 + 2) \\
		& \times\left| \frac{1}{2m_{D^*}} \left[ (m_{B}^2 - m_{D^*}^2 - q^2)(m_{B} + m_{D^*}) A_1(q^2)\right.\right.\\
		&\left.\left.-\frac{\lambda(m_{B}^2, m_{D^*}^2, q^2)}{m_{B} + m_{D^*}} A_2(q^2) \right] \right|^2 \Bigg\}, \\[2pt]
		\frac{d\Gamma^\pm(B \to D^* \ell\nu)}{dq^2} &= (1 - \hat{m}_\ell^2)^2 \frac{\sqrt{\lambda(m_{B}^2, m_{D^*}^2, q^2)} G_F^2 |V_{cb}|^2}{384 m_{B}^3 \pi^3} \Bigg\{ (m_\ell^2 + 2q^2)\lambda(m_{B}^2, m_{D^*}^2, q^2) \\
		& \times \left| \frac{V(q^2)}{m_{B} + m_{D^*}} \mp \frac{(m_{B} + m_{D^*}) A_1(q^2)}{\sqrt{\lambda(m_{B}^2, m_{D^*}^2, q^2)}} \right|^2 \Bigg\},
	\end{align*}
\end{footnotesize}
where $\lambda(q^2)=\lambda(m^{2}_{B},m^{2}_{D^*},q^{2})=(m^{2}_{B}+m^{2}_{D^*}-q^{2})^{2}-4m^{2}_{B}m^{2}_{D^*}$ and $\hat m^2_{\ell}=m^2_{\ell}/q^2$ with $m_{\ell}$ being the lepton mass. Even though the electron and muon masses are very small relative to the tau mass, we retain full lepton mass dependence throughout numerical calculations to quantify the mass effects.
The combined transverse and total differential decay widths are defined via
\be
\frac{d \Gamma_{T}}{d q^{2}}=\frac{d \Gamma_{+}}{d q^{2}}+\frac{d \Gamma_{-}}{d q^{2}}, \quad \frac{d \Gamma}{d q^{2}}=\frac{d \Gamma_{L}}{d q^{2}}+\frac{d \Gamma_{T}}{d q^{2}}.
\label{ptoV}
\en
For the $B\to D^*\ell\nu_\ell$ decays, one can define the longitudinal polarization fraction, 
\be
f_{L}=\frac{\Gamma_{L}}{\Gamma_{L}+\Gamma_{+}+\Gamma_{-}}. \label{eq:fl}
\en
The analytical expression of the forward-backward asymmetry for the semileptonic decays considered in this work is defined as \cite{Sakaki:2013bfa}
\be
\mathcal{A}_{FB} = \frac{\int^1_0 {d\Gamma \over dcos\theta} dcos\theta - \int^0_{-1} {d\Gamma \over dcos\theta} dcos\theta}
{\int^1_{-1} {d\Gamma \over dcos\theta} dcos\theta} = \frac{\int b_\theta(q^2) dq^2}{\Gamma_{B\to D^{(*)}\ell\nu_\ell}},\label{eq:AFB}
\en
where $\theta$ is the angle between the three-momenta of the lepton $\ell$ and the initial meson in the $\ell\nu_\ell$ rest frame. The function $b_{\theta}(q^2)$ represents the angular coefficient and can be written as \cite{Sakaki:2013bfa}
\be
b_{\theta}(q^2) &= &\frac{G_F^2 |V_{cb}|^2}{128\pi^3 m_{B}^3} \lambda(q^2) \left(1 - \frac{m_\ell^2}{q^2}\right)^2 \frac{m_\ell^2}{q^2} (m_{B}^2 - m_{D}^2)F_1(q^2)F_0(q^2), \label{fabyanb}
\en
for the semileptonic decays $B\to D\ell\nu_\ell$, and that for the semileptonic decays $B\to D^*\ell\nu_\ell$ is given as  
\be
b_{\theta}(q^2) &= &\frac{G_F^2 |V_{cb}|^2}{128\pi^3 m_{B_{(s)}}^3} q^2 \sqrt{\lambda(q^2)} \left(1 - \frac{m_\ell^2}{q^2}\right)^2 \left[ \frac{1}{2}\left(H_{V,+}^2 - H_{V,-}^2\right) + \frac{m_\ell^2}{q^2} (H_{V,0} H_{V,t})\right],\label{eq:BST}
\en
where the helicity amplitudes are defined as
\be
H_{V,\pm}(q^2) &=& (m_{B} + m_{D^*}) A_1(q^2) \mp \frac{\sqrt{\lambda(q^2)}}{m_{B} + m_{D^*}} V(q^2), \\
H_{V,0}(q^2) &=& \frac{m_{B} + m_{D^*}}{2m_{D^*} \sqrt{q^2}} \left[ -\left(m_{B}^2 - m_{D^*}^2 - q^2\right) A_1(q^2) + \frac{\lambda(q^2) A_2(q^2)}{(m_{B} + m_{D^*})^2} \right], \\
H_{V,t}(q^2) &= &-\sqrt{\frac{\lambda(q^2)}{q^2}} A_0(q^2).
\en
It is noticed that the subscript $V$ in each helicity amplitude refers to the $\gamma_\mu(1-\gamma_5)$ current.

\section{ NUMERICAL RESULTS OF Semi-leptonic Decays}\label{form2}
%\subsection{ Semi-leptonic decays}
Numerical inputs for meson and lepton masses, the CKM matrix elements, and $B_{(s)}$ meson lifetimes are taken from PDG \cite{ParticleDataGroup:2024cfk}. In this section, we compare our predictions for the branching ratios of the semi-leptonic decays $B_{(s)}\to D_{(s)}(1S,2S)\ell\nu_\ell$ and $B_{(s)}\to D^*_{(s)}(1S,2S)\ell\nu_\ell$ with those obtained from other theoretical methods and the available experimental data. The theoretical approaches employed in related research include the RQM \cite{Faustov:2022ybm,Faustov:2012mt,Bhol:2014jta}, the Bethe-Salpeter (BS) equation \cite{Zhou:2020ijj,Chen:2011ut,Wang:2012wk}, the covariant quark model (CQM) \cite{Ivanov:2015tru,Zhao:2006at,Azizi:2008vt,Azizi:2008tt}, the heavy quark effective theory (HQET) \cite{Fajfer:2012vx}, the perturbative QCD (PQCD) approach \cite{Hu:2019bdf,Fan:2015kna}, the
QCDSR \cite{Azizi:2008tt,Zhang:2021wnv}, the lattice QCD (LQCD) \cite{FermilabLattice:2015ilb,ECM,Harrison:2023dzh,Rahmani:2024pko,Dutta:2018jxz} and the LCSR \cite{Li:2009wq}. The uncertainties of our results arise from the lifetime of the $B_{(s)}$ meson, the decay constants of initial and final state mesons, respectively. The following are some comments:
\begin{table}[H]
	\caption{The branching ratios of the semi-leptonic decays $B\to D(1S) \ell^{+} \nu_{\ell} (\%)$ with the results from other theoretical approaches and data for comparison. }
	\begin{center}
		\scalebox{0.9}{
			\begin{tabular}{cccc}
				\hline\hline
			Decay Modes&$B^{0}\to D^{-}(1S) e^{+}{\nu}_{e}$&$B^{0}\to D^{-}(1S) \mu^{+}{\nu}_{\mu}$&$B^{0}\to D^{-}(1S) \tau^{+}{\nu}_{\tau}$\\
				\hline
				%This work(CLFQM)&$1.89^{+0.00+0.00+0.06}_{-0.00-0.06-0.06}$&$1.89^{+0.00+0.00+0.06}_{-0.00-0.06-0.06}$&$0.49^{+0.00+0.03+0.02}_{-0.00-0.01-0.01}$\\
				This work(CLFQM)&$1.96^{+0.01+0.00+0.06}_{-0.01-0.01-0.05}$&$1.96^{+0.01+0.00+0.06}_{-0.01-0.01-0.05}$&$0.51^{+0.00+0.01+0.02}_{-0.00-0.01-0.02}$\\
				%RQM\cite{Patnaik:2025fry}&-&-&0.75\\
				PQCD \cite{Hu:2019bdf}&$2.19$&$2.19$&$0.82$\\
				PQCD+Lattice \cite{Hu:2019bdf}&$1.95$&$1.95$&$0.62$\\
				CQM \cite{Ivanov:2015tru}&$2.74$&$2.74$&$0.73$\\
				HQET \cite{Fajfer:2012vx}&$-$&$-$&$0.64$\\
				PQCD \cite{Fan:2015kna}&$2.03$&$2.03$&$0.87$\\
				CPQCD \cite{Liu:2025ass}&$-$&$1.65$&$0.554$\\				PDG \cite{ParticleDataGroup:2024cfk}&$2.10\pm0.07$&$2.10\pm0.07$&$0.98\pm0.21$\\
				\hline
				Decay Modes&$B^{+}\to D^{0}(1S) e^{+}{\nu}_{e}$&$B^{+}\to D^{0}(1S) \mu^{+}{\nu}_{\mu}$&$B^{+}\to D^{0}(1S) \tau^{+}{\nu}_{\tau}$\\
				\hline
				%This work(CLFQM)&$2.04^{+0.00+0.00+0.06}_{-0.00-0.06-0.06}$&$2.04^{+0.00+0.01+0.06}_{-0.00-0.06-0.06}$&$0.53^{+0.00+0.03+0.02}_{-0.00-0.02-0.02}$\\
				This work(CLFQM)&$2.11^{+0.01+0.00+0.06}_{-0.01-0.01-0.06}$&$2.11^{+0.01+0.00+0.06}_{-0.01-0.01-0.06}$&$0.55^{+0.00+0.01+0.02}_{-0.00-0.01-0.02}$\\
				RQM \cite{Faustov:2022ybm}&$2.53$&$2.53$&$0.68$\\
				%RQM\cite{Patnaik:2025fry}&-&-&0.81\\
				PQCD \cite{Hu:2019bdf}&$2.29$&$2.29$&$0.86$\\
				PQCD+Lattice \cite{Hu:2019bdf}&$2.10$&$2.10$&$0.69$\\
				LQCD \cite{FermilabLattice:2015ilb,ECM,Harrison:2023dzh}&$-$&$-$&$0.65$\\
				HQET \cite{Fajfer:2012vx}&$-$&$-$&$0.66$\\
				PQCD \cite{Fan:2015kna}&$2.19$&$2.19$&$0.95$\\
				PDG \cite{ParticleDataGroup:2024cfk}&$2.26\pm0.07$&$2.26\pm0.07$&$0.77\pm0.25$\\
				\hline
		\end{tabular}}\label{table1}
	\end{center}
\end{table}
Our predictions for the branching ratios of the semi-leptonic decays $B\to D(1S) \ell \nu_{\ell}$ are listed in Table \ref{table1}. It is noted that the branching ratios of the decays $B\to D(1S) \ell^{\prime} \nu_{\ell^{\prime}}$ are in good agreement with the PQCD+Lattice results \cite{Hu:2019bdf} and deviate from the experimental data by about $1.5\sigma$. While for the decay $B^0\to D^-(1S) \tau^{+} \nu_{\tau}$, the tension reaches $2.2\sigma$. Although the deviation for the decay $B^+\to D^0(1S) \tau^{+} \nu_{\tau}$ is milder at only $0.9\sigma$, the current experimental errors remain sizable. An analogous condition holds for the decays $B\to D^*(1S) \ell \nu_{\ell}$ as shown in Table \ref{b3}. The branching ratios of the decays $B\to D^{*}(1S)\ell^\prime\nu_{\ell^\prime}$ and $B_s\to D^{*}_s(1S)\mu\nu_{\mu}$ agree well with the data with only a $0.3\sigma$ deviation. However, the branching ratios of the decays $B\to D^*(1S) \tau\nu_{\tau}$ show a deviation exceeding $3\sigma$ from experimental data. In a word, our predictions for the decays $B\to D^{(*)}(1S)\ell^{\prime}\nu_{\ell^\prime}$ can explain the data well, while the branching ratios of the decays $B\to D^{(*)}(1S)\tau{\nu_\tau}$ are systematically smaller than the data. The similar situations also exist in other theoretical results. By the way, it is contrary for the nonleptonic B decays, such as $B\to D(1S)\pi(K, \rho, K^*)$, where the theoretical results are larger than the data.
\begin{table}[H]
	\caption{The branching ratios of the semi-leptonic decays $B\to D^*(1S) l^{+} \nu_{l}(\%)$ with the results from other theoretical approaches and data for comparison.}
	\begin{center}
		\scalebox{1}{
			\begin{tabular}{cccc}
				\hline\hline
				Decay Modes&$B^{0}\to D^{*-}(1S) e^{+}{\nu}_{e}$&$B^{0}\to D^{*-}(1S) \mu^{+}{\nu}_{\mu}$&$B^{0}\to D^{*-}(1S) \tau^{+}{\nu}_{\tau}$\\
				\hline
				This work&$4.97^{+0.01+0.03+0.36}_{-0.01-0.01-0.16}$&$4.95^{+0.01+0.03+0.36}_{-0.01-0.01-0.16}$&$1.13^{+0.00+0.01+0.07}_{-0.00-0.01-0.04}$\\
				RQM \cite{Faustov:2022ybm}	&$6.28$&$6.28$&$1.45$\\
				PQCD \cite{Hu:2019bdf}&$5.32$&$5.32$&$1.53$\\
				PQCD+Lattice \cite{Hu:2019bdf}&$4.63$&$4.63$&$1.25$\\
				PQCD \cite{Fan:2015kna}&$4.52$&$4.52$&$1.36$\\
				HQET \cite{Fajfer:2012vx}&$-$&$-$&$1.29$\\
				CQM \cite{Ivanov:2015tru}&$6.64$&$6.64$&$1.57$\\
				CPQCD \cite{Liu:2025ass}&$-$&$4.33$&$1.175$\\				PDG \cite{ParticleDataGroup:2024cfk}&$4.87\pm0.09$&$4.87\pm0.09$&$1.48\pm0.09$\\
				\hline
				Decay Modes&$B^{+}\to D^{*0}(1S) e^{+}{\nu}_{e}$&$B^{+}\to D^{*0}(1S) \mu^{+}{\nu}_{\mu}$&$B^{+}\to D^{*0}(1S) \tau^{+}{\nu}_{\tau}$\\
				\hline
				This work&$5.37^{+0.01+0.03+0.39}_{-0.01-0.01-0.17}$&$5.35^{+0.01+0.03+0.38}_{-0.01-0.01-0.17}$&$1.22^{+0.00+0.01+0.08}_{-0.00-0.01-0.04}$\\RQM\cite{Faustov:2022ybm}	&$6.81$&$6.77$&$1.52$\\
				PQCD \cite{Hu:2019bdf}&$5.53$&$5.53$&$1.60$\\
				PQCD+Lattice \cite{Hu:2019bdf}&$4.89$&$4.89$&$1.34$\\
				PQCD \cite{Fan:2015kna}&$4.87$&$4.87$&$1.47$\\
				HQET \cite{Fajfer:2012vx}&$-$&$-$&$1.34$\\
				PDG \cite{ParticleDataGroup:2024cfk}	&$5.26$&$5.26$&$1.88$\\
				\hline
			\end{tabular}\label{b3}}
	\end{center}
\end{table}

Tensions between the SM expectations and experimental measurements have driven extensive theoretical studies in the search for NP effects.
Before claiming a new physics signal, it is crucial to subtantially boost the precision of both theoretical calculations and experimental measurements. Current experiments still have relatively large uncertainties in the measured branching ratios of decays $B\to D(1S) \tau \nu_{\tau}$ shown in Table \ref{table1}. 

%The similar trend is also observed in the decays $B\to D^{*}(1S) \tau\nu_{\tau}$, that is most theoretical predictions for their branching raitos are also smaller than the data as shown in Table \ref{b3}. While our predictions for the branching ratios of the decays $B\to D^{*}(1S) \ell^\prime\nu_{\ell^\prime}$ are larger than the data by about $20\%$. Maybe it is important to precisely determine the decay constant of $D^*{*}$ meson.

For the branching ratios of the semi-leptonic decays $B_s \to D_s(1S)\mu\nu_\mu$ and $B_s \to D_s^{*}(1S)\mu\nu_\mu$, the discrepancies between our predictions and the data are approximately $1.1\sigma$ and $0.3\sigma$, respectively.  From Table \ref{b6}, one can find that the branching ratios of the decays $B_s \to D_s\ell^\prime\nu_{\ell^\prime}$ given by the LCSR \cite{Zhang:2022opp,Li:2009wq}, BS equation \cite{Chen:2011ut}, and PQCD \cite{Hu:2019bdf} approaches are smaller than $2\times10^{-2}$. In contrast, those from the QCDSR\cite{Azizi:2008tt}, CQM \cite{Zhao:2006at,Azizi:2008vt,Soni:2021fky} and RQM \cite{Bhol:2014jta} are larger than $2.5\times10^{-2}$. Meanwhile, the results obtained by the same methods as listed above, such as the QCDSR \cite{Zhang:2021wnv}, PQCD \cite{Fan:2013kqa} and RQM \cite{Faustov:2012mt}, fall within the range $(2.013-2.469)\times10^{-2}$ given by the LQCD \cite{Dutta:2018jxz} and well account for the data. In Table \ref{b7}, apart for the results given by the PQCD+Lattice \cite{Hu:2019bdf} and  CQM \cite{Soni:2021fky,Zhao:2006at}, many other theoretical predictions are consistent with the available data within errors. Further precise experimental data are needed, which are useful not only
 for clarifying different theoretical approaches but also for constraining parameters in relevant models.
\begin{table}[H]
	\caption{ The branching ratios of the semi-leptonic decays $B_{s}^{0}\to D_{s}^{-}(1S) l^{+} \nu_{l}$ $(\%)$ with the results from other theoretical approaches and data for comparison.}
	\begin{center}
		\scalebox{1}{
			\begin{tabular}{cccc}
				\hline\hline
				Decay Modes&$B_{s}^{0}\to D_{s}^{-}(1S) e^{+}{\nu}_{e}$&$B_{s}^{0}\to D_{s}^{-}(1S) \mu^{+}{\nu}_{\mu}$&$B_{s}^{0}\to D_{s}^{-}(1S) \tau^{+}{\nu}_{\tau}$\\
				\hline
				%This work(CLFQM)&$1.97^{+0.01+0.07+0.00}_{-0.01-0.06-0.01}$&$1.97^{+0.01+0.07+0.00}_{-0.01-0.06-0.01}$&$0.51^{+0.00+0.03+0.00}_{-0.00-0.00-0.03}$\\
				This work(CLFQM)&$2.05^{+0.00+0.11+0.05}_{-0.01-0.11-0.06}$&$2.04^{+0.01+0.11+0.01}_{-0.01-0.11-0.01}$&$0.53^{+0.00+0.03+0.03}_{-0.00-0.07-0.02}$\\
				NCQM \cite{Albertus:2014eqa}&$2.32$&$-$&$0.67$\\
				LCSR \cite{Zhang:2022opp}&$1.817$&$1.817$&$0.606$\\
				QCDSR \cite{Azizi:2008tt}&$2.8-3.5$&$2.8-3.5$&$-$\\
				QCDSR \cite{Zhang:2021wnv}&$2.03$&$2.03$&$-$\\
				LCSR \cite{Li:2009wq}&$1.0$&$1.0$&$0.33$\\
				PQCD \cite{Fan:2013kqa}&$2.13$&$2.13$&$0.84$\\
				RQM \cite{Faustov:2012mt}&$2.1$&$2.1$&$0.62$\\
				RQM \cite{Bhol:2014jta}&$2.54$&$2.54$&$0.70$\\
				CQM \cite{Zhao:2006at}&$2.73-3.00$&$2.73-3.00$&$-$\\
				PQCD \cite{Hu:2019bdf}&$1.97$&$1.97$&$0.72$\\
				PQCD+Lattice \cite{Hu:2019bdf}&$1.84$&$1.84$&$0.63$\\
				LQCD \cite{Dutta:2018jxz}&$2.013-2.469$&$2.013-2.469$&$0.619-0.724$\\
				BS \cite{Chen:2011ut}&$1.4-1.7$&$1.4-1.7$&$0.47-0.55$\\
				LQCD \cite{Rahmani:2024pko}&$2.31$&$2.31$&$0.69$\\
				CQM \cite{Azizi:2008vt,Azizi:2008tt}&$2.8-3.8$&$2.8-3.8$&$-$\\
				CQM \cite{Soni:2021fky}&$2.89$&$2.88$&$0.78$\\		
                PDG \cite{ParticleDataGroup:2024cfk}&$-$&$2.29\pm0.21$&$-$\\
				\hline
			\end{tabular}\label{b6}}
	\end{center}
\end{table}

Compared to the semi-leptonic decays $B\to D^{(*)}(1S)\ell\nu_\ell$, the theoretical studies on the semi-leptonic $B\to D^{(*)}(2S)\ell\nu_\ell$ decays remain relatively scarce. Currently, only the BS equation method \cite{Zhou:2020ijj,Wang:2012wk} and the RQM \cite{Faustov:2012mt}
have been applied to in this class of decays. In Tables \ref{tb5} and \ref{b8}, we present the branching ratios of the semileptonic decays $B\to D^{(*)}(2S)\ell\nu_\ell$ and $B_s\to D^{(*)}_s(2S)\ell\nu_\ell$, respectively, which are at least one order of magnitude smaller than those of the decays $B\to D^{(*)}(1S)\ell\nu_\ell$ and $B_s\to D^{(*)}_s(1S)\ell\nu_\ell$. These semileptonic $B_{(s)}$ decays to the first radially excited charmed mesons have branching ratios up to $10^{-4}\sim10^{-3}$, which are within the detection accuracy of current experiments. It is noted that our predictions are much larger than those from the BS equation approach \cite{Zhou:2020ijj}, while being consistent with the RQM results \cite{Faustov:2012mt}. A similar hierarchical pattern is also observed in nonleptonic $B_{(s)}$ meson decays as documented in our prior work \cite{Sun:2025yiz}. 
\begin{table}[H]
	\caption{The branching ratios of the semi-leptonic decays $B_{s}^{0}\to D_{s}^{*-}(1S) l^{+} \nu_{l}$ $(\%)$ with the results from other theoretical approaches and data for comparison.   }
	\begin{center}
		\scalebox{1}{
			\begin{tabular}{cccc}
				\hline\hline
				Decay Modes&$B_{s}^{0}\to D_{s}^{*-}(1S) e^{+}{\nu}_{e}$&$B_{s}^{0}\to D_{s}^{*-}(1S) \mu^{+}{\nu}_{\mu}$&$B_{s}^{0}\to D_{s}^{*-}(1S) \tau^{+}{\nu}_{\tau}$\\
				\hline
				%This work(CLFQM)&$6.17^{+0.02+0.25+0.41}_{-0.02-0.27-0.47}$&$6.15^{+0.02+0.25+0.41}_{-0.02-0.27-0.46}$&$1.38^{+0.00+0.05+0.08}_{-0.00-0.06-0.09}$\\
				This work(CLFQM)&$5.02^{+0.02+0.25+0.41}_{-0.02-0.27-0.47}$&$5.00^{+0.02+0.25+0.41}_{-0.02-0.27-0.46}$&$1.13^{+0.00+0.05+0.08}_{-0.00-0.06-0.09}$\\
				NCQM\cite{Albertus:2014eqa}&$6.26$&$-$&$1.53$\\
				PQCD\cite{Fan:2013kqa}&$4.76$&$4.76$&$1.44$\\
				RQM\cite{Faustov:2012mt}&$5.3$&$5.3$&$1.3$\\
				RQM\cite{Bhol:2014jta}&$2.54$&$2.54$&$0.70$\\
				CQM\cite{Zhao:2006at}&$7.49-7.66$&$7.49-7.66$&$-$\\
				PQCD\cite{Hu:2019bdf}&$5.04$&$5.04$&$1.45$\\
				PQCD+Lattice\cite{Hu:2019bdf}&$4.42$&$4.42$&$1.20$\\
				BS\cite{Chen:2011ut}&$5.1-5.8$&$5.1-5.8$&$1.2-1.3$\\
				LQCD\cite{Rahmani:2024pko}&$5.25$&$5.25$&$1.31$\\
				CQM\cite{Azizi:2008vt,Azizi:2008tt}&$1.89-6.61$&$1.89-6.61$&$-$\\
				CQM\cite{Soni:2021fky}&$6.42$&$6.39$&$1.53$\\	PDG \cite{ParticleDataGroup:2024cfk}&$-$&$5.2\pm0.5$&$-$\\
				\hline
			\end{tabular}\label{b7}}
	\end{center}
\end{table}

\begin{table}[H]
	\caption{ The branching ratios of the semi-leptonic decays $B\to D^{(*)}(2S)  l^{+} \nu_{l}$ $(10^{-3})$ with the results from the BS equation for comparison. }
	\begin{center}
		\scalebox{1}{
			\begin{tabular}{ccc|ccc}
				\hline\hline
				Decay Modes&This work&BS\cite{Zhou:2020ijj}&Decay Modes&This work&BS\cite{Zhou:2020ijj}\\
				\hline
				$B^{0}\to D^{-}(2S) e^{+}{\nu}_{e}$&$1.44^{+0.00+0.18+0.15}_{-0.00-0.19-0.19}$&$0.139$&$B^{+}\to D^{0}(2S) e^{+}{\nu}_{e}$&$1.55^{+0.00+0.19+0.17}_{-0.00-0.21-0.20}$&$0.15$\\
				$B^{0}\to D^{-}(2S) \mu^{+}{\nu}_{\mu}$&$1.43^{+0.00+0.18+0.15}_{-0.00-0.19-0.19}$&$0.138$&$B^{+}\to D^{0}(2S) \mu^{+}{\nu}_{\mu}$&$1.54^{+0.00+0.19+0.17}_{-0.00-0.21-0.20}$&$0.149$\\
				$B^{0}\to D^{-}(2S) \tau^{+}{\nu}_{\tau}$&$0.18^{+0.00+0.03+0.03}_{-0.00-0.03-0.03}$&$0.0135$&$B^{+}\to D^{0}(2S) \tau^{+}{\nu}_{\tau}$&$0.19^{+0.00+0.03+0.03}_{-0.00-0.04-0.04}$&$0.0149$\\
				$B^{0}\to D^{*-}(2S) e^{+}{\nu}_{e}$&$2.67^{+0.00+0.20+0.11}_{-0.00-0.24-0.14}$&$0.1942$&$B^{+}\to D^{*0}(2S) e^{+}{\nu}_{e}$&$2.93^{+0.00+0.21+0.13}_{-0.00-0.26-0.16}$&$0.2123$\\
				$B^{0}\to D^{*-}(2S) \mu^{+}{\nu}_{\mu}$&$2.65^{+0.00+0.20+0.11}_{-0.00-0.24-0.14}$&$0.1932$&$B^{+}\to D^{*0}(2S) \mu^{+}{\nu}_{\mu}$&$2.91^{+0.00+0.21+0.13}_{-0.00-0.26-0.16}$&$0.2113$\\
				$B^{0}\to D^{*-}(2S) \tau^{+}{\nu}_{\tau}$&$0.19^{+0.00+0.02+0.10}_{-0.00-0.02-0.11}$&$0.0137$&$B^{+}\to D^{*0}(2S) \tau^{+}{\nu}_{\tau}$&$0.21^{+0.00+0.02+0.11}_{-0.00-0.03-0.11}$&$0.0155$\\
				\hline
			\end{tabular}\label{tb5}}
	\end{center}
\end{table}
\begin{table}[H]
	\caption{The branching ratios of the semi-leptonic decays $B_{s}^{0}\to D_{s}^{(*)-}(2S) l^{+} \nu_{l}$ $(10^{-3})$ with the results from the BS equation and RQM for comparison.  }
	\begin{center}
		\scalebox{0.8}{
			\begin{tabular}{ccccc|ccccc}
				\hline\hline
				Decay Modes&This work&BS\cite{Zhou:2020ijj}&BS\cite{Wang:2012wk}&RQM\cite{Faustov:2012mt}&Decay Modes&This work&\cite{Zhou:2020ijj}&RQM\cite{Faustov:2012mt}\\
				\hline
				$B_{s}^{0}\to D_{s}^{-}(2S) e^{+}{\nu}_{e}$&$1.35^{+0.00+0.25+0.14}_{-0.00-0.23-0.22}$&$0.314$&$0.99$&$2.7$&$B_{s}^{0}\to D_{s}^{*-}(2S) e^{+}{\nu}_{e}$&$2.25^{+0.01+1.09+0.64}_{-0.01-0.70-0.80}$&$0.5873$&$3.8$\\
				$B_{s}^{0}\to D_{s}^{-}(2S) \mu^{+}{\nu}_{\mu}$&$1.34^{+0.00+0.25+0.14}_{-0.00-0.23-0.22}$&$0.312$&$-$&$-$&$B_{s}^{0}\to D_{s}^{*-}(2S) \mu^{+}{\nu}_{\mu}$&$2.23^{+0.01+1.09+0.64}_{-0.01-0.70-0.80}$&$0.5842$&$-$\\
				$B_{s}^{0}\to D_{s}^{-}(2S) \tau^{+}{\nu}_{\tau}$&$0.14^{+0.00+0.03+0.02}_{-0.00-0.03-0.03}$&$0.0244$&$-$&$0.11$&$B_{s}^{0}\to D_{s}^{*-}(2S) \tau^{+}{\nu}_{\tau}$&$0.15^{+0.00+0.05+0.06}_{-0.00-0.07-0.06}$&$0.0405$&$0.15$\\
				\hline
			\end{tabular}\label{b8}}
	\end{center}
\end{table}
Both semi-leptonic decays $B\to D^{(*)}(1S)\ell\nu_\ell$ and $B\to D^{(*)}(2S)\ell\nu_\ell$ are induced by the same $b\to c\ell\nu_\ell$ transition, so the decays $B\to D^{(*)}(2S)\ell\nu_\ell$ can also serve as a probe for NP searches. The
tensions between experimental data and SM predictions in the $B\to D^{(*)}(1S)\ell\nu_\ell$ decays have attracted widespread attention. Whether a similar anomaly also occurs in the semi-leptonic $B\to D^{(*)}(2S)\ell\nu_\ell$ decays is worthy of in-depth investigation. There is a long-standing puzzle in flavor physics about determining the $V_{cb}$, that is $3\sigma$ deviations between the inclusive and 
exclusive determinations. To resolve this inconsistency, the LHCb collaboration has determined the following branching fraction ratios \cite{lhcb1},
\be
	\frac{\mathcal{B}r(B_s^0 \to D_s^- \mu^+ \nu_\mu)}{\mathcal{B}r(B_s^0 \to D_s^{*-} \mu^+ \nu_\mu)} &=&0.464 \pm 0.013_{\text{stat}} \pm 0.043_{\text{syst}} \pm 0.05_{\text{ext}},\\
	\frac{\mathcal{B}r(B_s^0 \to D_s^- \mu^+ \nu_\mu)}{\mathcal{B}r(B^0 \to D^- \mu^+ \nu_\mu)} &=& 1.09 \pm 0.05_{\text{stat}} \pm 0.06_{\text{syst}} \pm 0.05_{\text{ext}},\\
	\frac{\mathcal{B}r(B_s^0 \to D_s^{*-} \mu^+ \nu_\mu)}{\mathcal{B}r(B^0 \to D^{*-} \mu^+ \nu_\mu)} &=& 1.06 \pm 0.05_{\text{stat}} \pm 0.07_{\text{syst}} \pm 0.05_{\text{ext}}.
\en
 For comparison, we present our predictions for the corresponding branching fraction ratios, as well as those for the case where $D_{(s)}(1S)$ and $D^{*}_{(s)}(1S)$ are replaced by $D_{(s)}(2S)$ and $D^{*}_{(s)}(2S)$, respectively, 
\be
	\frac{\mathcal{B}r(B_s^0 \to D_s^- \mu^+ \nu_\mu)}{\mathcal{B}r(B_s^0 \to D_s^{*-} \mu^+ \nu_\mu)} &=& 0.408\pm0.029 \;\;\;\;\;\; \frac{\mathcal{B}r(B_s^0 \to D_s^- (2S)\mu^+ \nu_\mu)}{\mathcal{B}r(B_s^0 \to D_s^{*-}(2S) \mu^+ \nu_\mu)} = 0.601\pm0.342,\\
	\frac{\mathcal{B}r(B_s^0 \to D_s^- \mu^+ \nu_\mu)}{\mathcal{B}r(B^0 \to D^- \mu^+ \nu_\mu)} &=& 1.041\pm0.074  \;\;\;\;\;\; \frac{\mathcal{B}r(B_s^0 \to D_s^-(2S) \mu^+ \nu_\mu)}{\mathcal{B}r(B^0 \to D^-(2S) \mu^+ \nu_\mu)} = 0.937\pm0.269,\\
	\frac{\mathcal{B}r(B_s^0 \to D_s^{*-} \mu^+ \nu_\mu)}{\mathcal{B}r(B^0 \to D^{*-} \mu^+ \nu_\mu)} &=& 1.010\pm0.216 \;\;\;\;\;\; \frac{\mathcal{B}r(B_s^0 \to D_s^{*-}(2S) \mu^+ \nu_\mu)}{\mathcal{B}r(B^0 \to D^{*-}(2S) \mu^+ \nu_\mu)} = 0.842\pm0.446,
\en
where the errors stem from the $B_{(s)}$ meson lifetime, the decay
constants of initial and final state mesons. 
%As is widely recognized in the field, the Standard Model (SM) is not a complete theoretical framework, particularly at elevated energy scales. It is therefore of crucial significance to carry out high-precision tests of the SM, with the aim of probing for signals of new physics (NP) beyond the SM\cite{liyin}. Recently, a number of experimental collaborations have reported a series of anomalous measurements of the ratios $\mathcal{R}(D_{(s)}^{\ast})$, which are defined as(1910.06595) 

As one of the pillars of the SM, the ratios $\mathcal{R}(D^{(\ast)})$ and $\mathcal{R}(D_{s}^{(\ast)})$ are a powerful test of the LFU. Their values have attracted widespread attention from both experientalists and theorists. Our predictions are presented in Table \ref{bR}, along with experimental measurements and other theoretical results for comparison.
\begin{table}[H]
	\caption{The values of ratios $\mathcal{R}(D^{(\ast)})$ and $\mathcal{R}(D_{s}^{(\ast)})$, where the individual errors, as those in branching ratios, have been added in quadrature. }
	\begin{center}
		\scalebox{1}{
			\begin{tabular}{ccccc}
				\hline\hline
				&$\mathcal{R}\left(D^{-}(1S)\right)$&$\mathcal{R}\left(D^{0}(1S)\right)$&$\mathcal{R}\left(D^{*-}(1S)\right)$
				&$\mathcal{R}\left(D^{*0}(1S)\right)$\\
				\hline
This work&$0.260\pm0.014$&$0.261\pm0.013$&$0.228\pm0.026$&$0.228\pm0.026$\\
RQM \cite{Faustov:2022ybm}&$-$&$0.269$&$0.231$&$0.224$\\
PQCD+Lattice \cite{Hu:2019bdf}&$0.318$&$0.329$&$0.270$&$0.274$\\
PQCD \cite{Fan:2015kna}&$0.429$&$0.434$&$0.301$&$0.302$\\
Belle-II \cite{Belle-II:2025yjp}&$0.418$&$-$&$0.306$&$-$\\
LHCb \cite{RA2025} &$0.249$&$-$&$0.402$&$-$\\          
\hline
&$\mathcal{R}\left(D^{-}(2S)\right)$&$\mathcal{R}\left(D^{0}(2S)\right)$&$\mathcal{R}\left(D^{*-}(2S)\right)$
&$\mathcal{R}\left(D^{*0}(2S)\right)$\\\hline
This work&$0.125\pm0.037$&$0.123\pm0.039$&$0.072\pm0.0431$&$0.072\pm0.043$\\
BS \cite{Zhou:2020ijj}&$0.097$&$0.099$&$0.071$&$0.073$\\ \hline
&$\mathcal{R}\left(D_{s}^{-}(1S)\right)$&$\mathcal{R}\left(D_{s}^{*-}(1S)\right)$&$\mathcal{R}\left(D_{s}^{-}(2S)\right)$
&$\mathcal{R}\left(D_{s}^{*-}(2S)\right)$\\ \hline
This work&$0.259\pm0.033$&$0.226\pm0.031$&$0.104\pm0.038$&$0.067\pm0.052$\\ 
RQM \cite{Bhol:2014jta}&$0.276$&$0.276$&$-$&$-$\\            
RQM \cite{Faustov:2012mt}&$0.295$&$0.245$&$-$&$-$\\
PQCD+Lattice \cite{Hu:2019bdf} &$0.342$&$0.271$&$-$&$-$\\
BS \cite{Zhou:2020ijj}&$-$&$-$ &$0.078$&$0.069$\\          				
				\hline\hline
		\end{tabular}\label{bR}}
	\end{center}
\end{table}
From the initial value $0.44\pm0.058\pm0.042$ reported by Belle \cite{BaBar:2012obs} to the recent one $0.249\pm0.043\pm0.047$ given by LHCb, the $\mathcal{R}_D$ measurements exhibit a decreasing trend over time. Accordingly, relative to the experimental measurements, our prediction $\mathcal{R}\left(D\right)=0.260\pm0.014$ yields a shrinking tension, which changes from $2.4\sigma$ to $0.16\sigma$, while the deviation compared to the HFAG value still amounts to $3.1\sigma$. It is similar for the $\mathcal{R}(D^*)$ values between our predictions and the data. For example, our value $\mathcal{R}(D^{*-})=0.228\pm0.026$ compares with the measurement $0.260\pm0.015\pm0.016\pm0.012$ given by LHCb \cite{LHCb:2023uiv}, corresponding to a $1.3\sigma$ deviation, while sits at a $3.7\sigma$ below the HFAG average $0.295\pm0.010\pm0.010$. If we adopt the latest LHCb measurement $\mathcal{R}(D^*)=0.402\pm0.081\pm0.085$, the tension reduces to $1.5\sigma$, which falls within the range allowed by statistical fluctuations again. More precise experimental data for $\mathcal{R}(D_{(s)})$ and $\mathcal{R}(D^*_{(s)})$ are urgently needed. Certainly, our predictions are consistent with the results obtained from the RQM \cite{Faustov:2022ybm,Faustov:2012mt,Bhol:2014jta} and the PQCD\cite{Fan:2015kna} calculations within errors.
%\subsection{Other physical observables}

Next, we calculate other two physical observables, namely the forward-backward asymmetry $\mathcal{A}_{\text{FB}}$ and the longitudinal polarization fraction $f_{L}$, using Eqs. (\ref{eq:fl}) and (\ref{eq:AFB}). Their values are listed in Tables \ref{AFB1}, \ref{AFB2} and \ref{FLE}.  These physical quantities are also sensitive to some kinds of new physics and can be measured in the present experiments. So they have attracted widespread attention from various theoretical approaches.
From our calculations, we arrive at the following conclusions. 
 \begin{itemize}
 	\item 
 	The forward-backward asymmetries $\mathcal{A}_{\text{FB}}$ for the decays $B_{(s)} \to D_{(s)} \ell \nu_\ell$ exhibit a clear hierarchical relationship, because they are proportional to the square of the lepton mass as shown in Eq. (\ref{fabyanb}). While the difference among $\mathcal{A}_{\text{FB}}$ values of the decays  $B_{(s)} \to D^*_{(s)} \ell \nu_\ell$ is not significant.
 	\item
 The $\mathcal{A}_{\text{FB}}$ values for the decays $B_{(s)} \to D_{(s)} \ell \nu_\ell$ are positive, while those of the decays $B_{(s)} \to D^*_{(s)} \ell \nu_\ell$ are negative. Our predictions are consistent with the results given by the PQCD and PQCD+Lattice approaches\cite{Hu:2019bdf} and HQET \cite{Huang:2018nnq}, while have opposite signs relative to the CQM \cite{Ivanov:2015tru} and RQM \cite{Faustov:2022ybm} calculations
  except for those of the decays $B_{(s)} \to D^*_{(s)} \tau \nu_\tau$. 
\item 
 It is worth noting that the magnitudes of the  $\mathcal{A}_{\text{FB}}$ values for the decays $B_{(s)} \to D_{(s)} \tau \nu_\tau$ given by these theoretical approaches are in excellent agreement with each other. However, the discrepancies among the results for the decays $B_{(s)} \to D_{(s)}^{*} \tau\nu_\tau$ are significant.
\item
From Table \ref{AFB2}, one can find that the forward-backward asymmetries of the decays $B_{(s)} \to D_{(s)}(2S)\ell \nu_\ell, D^*_{(s)}(2S)\ell \nu_\ell$ are similar to those of corresponding decays $B_{(s)} \to D_{(s)}(1S)\ell \nu_\ell,D^*_{(s)}(1S)\ell \nu_\ell$. These results can be verified by future LHCb and Belle II experiments. 
\end{itemize}

In order to investigate the dependence of the polarization on $q^2$, we calculate the longitudinal polarization fractions by dividing the full energy region into two regions for each decay as listed in Table \ref{FLE}, where Region 1 is defined as $m_{\ell}^2 < q^2 < \frac{(m_B-m_{D^*(nS)})^2+m_{\ell}^2}{2}$ and Region 2 is $\frac{(m_B-m_{D^*(nS)})^2+m_{\ell}^2}{2} < q^2 < (m_{B} - m_{D^*(nS)})^2$ with $n = 1,2$.
Some comments are in order.
For the decays $B_{(s)}\to D^{*}_{(s)}(nS)\ell^{\prime}\nu_{\ell^{\prime}}$, the longitudinal polarization fractions from Region 1 are larger than those from Region 2, while it is contrary to the decays$B_{(s)}\to D^{*}_{(s)}(nS)\tau\nu_{\tau}$.
For the decays $B_{(s)}\to D^{*}_{(s)}(nS)\ell^{\prime}\nu_{\ell^{\prime}}$, the longitudinal polarizations are larger than the transverse ones, while for the decays $B_{(s)}\to D^{*}_{(s)}(nS)\tau\nu_{\tau}$, the longitudinal polarization fractions are almost equal to, or even smaller than, the transverse ones. In other words, the longitudinal polarization fractions of the decays $B_{(s)}\to D^{*}_{(s)}(nS)\ell^{\prime}\nu_{\ell^{\prime}}$ are always larger than those of the decays $B_{(s)}\to D^{*}_{(s)}(nS)\tau\nu_{\tau}$. Furthermore, the longitudinal polarization fractions of the decays $B_{(s)}\to D^{*}_{(s)}(2S)\ell\nu_{\ell}$ are larger than those of the corresponding decays $B_{(s)}\to D^{*}_{(s)}(1S)\ell\nu_{\ell}$. These relations can be expressed using the following formulas 
\be
f_L(B_{(s)}\to D^{*}_{(s)}(nS)\ell^{\prime}\nu_{\ell^{\prime}})&>&f_L(B_{(s)}\to D^{*}_{(s)}(nS)\tau\nu_{\tau}), n=1,2,\\
f_L(B_{(s)}\to D^{*}_{(s)}(2S)\ell\nu_{\ell})&>&f_L(B_{(s)}\to D^{*}_{(s)}(1S)\ell\nu_{\ell}).
\en
All the results for our considered decays are consistent with other theoretical calculations, such as the RQM \cite{Faustov:2022ybm}, PQCD +Lattice\cite{Hu:2019bdf}, HQET \cite{Huang:2018nnq} and model independent analysis (MIA) \cite{Bhattacharya:2018kig}. 
It is worth noting that our prediction for the value of $f_L(B^0\to D^{*-}(1S)\tau^+\nu_{\tau})$ is in good agreement with the experimental data given by LHCb \cite{LHCb:2023ssl}, but smaller than Belle measurement \cite{Belle:2019ewo}. We look forward to more experimental data to verify our predictions.

\begin{table}[H]
	\caption{Forward-backward asymmetries $\mathcal{A}_{FB}$ for the decays $B_{(s)} \to D_{(s)}(1S)\ell\nu_{\ell}$ and $B_{(s)} \to D^{*}_{(s)}(1S)\ell\nu_{\ell}$ with the results from other theoretical approaches for comparison.}
	\begin{center}
		\scalebox{0.8}{
			\begin{tabular}{|c|c|c|c|}
				\hline\hline
				Channels  &$B^{0}\to D^{-}(1S) e^{+}{\nu}_{e}$&$B^{0}\to D^{-}(1S) \mu^{+}{\nu}_{\mu}$&$B^{0}\to D^{-}(1S) \tau^{+}{\nu}_{\tau}$\\
				\hline
				$\mathcal{A}_{FB}$&$4.36^{+0.01+0.01+0.13}_{-0.01-0.01-0.13}\times10^{-7}$&$0.02^{+0.00+0.00+0.00}_{-0.00-0.00-0.00}$&$0.36^{+0.00+0.00+0.01}_{-0.00-0.01-0.01}$\\
				CQM\cite{Ivanov:2015tru}&$-11.7\times10^{-7}$&$-$&$-0.36$\\
				PQCD\cite{Hu:2019bdf}&$-$&$-$&$0.35$\\
				PQCD+Lattice\cite{Hu:2019bdf}&$-$&$-$&$0.36$\\
				\hline
				Channels&$B^{+}\to D^{0}(1S)e^{+}\nu_{e}$&$B^{+}\to D^{0}(1S)\mu^{+}\nu_{\mu}$&$B^{+}\to D^{0}(1S)\tau^{+}\nu_{\tau}$\\
				\hline
				$\mathcal{A}_{FB}$&$4.36^{+0.01+0.01+0.13}_{-0.01-0.01-0.13}\times10^{-7}$&$0.02^{+0.00+0.00+0.00}_{-0.00-0.00-0.00}$&$0.36^{+0.00+0.04+0.01}_{-0.00-0.05-0.01}$\\
				RQM\cite{Faustov:2022ybm}&$-9.8\times10^{-7}$&$-0.013$&$-0.37$\\
				\hline
				Channels&$B_{s}^{0}\to D_{s}^{-}(1S)e^{+}\nu_{e}$&$B_{s}^{0}\to D_{s}^{-}(1S)\mu^{+}\nu_{\mu}$&$B_{s}^{0}\to D_{s}^{-}(1S)\tau^{+}\nu_{\tau}$\\
\hline
$\mathcal{A}_{FB}$&$4.39^{+0.06+0.26+0.14}_{-0.06-0.19-0.13}\times10^{-7}$&$0.02^{+0.00+0.00+0.00}_{-0.00-0.00-0.00}$&$0.36^{+0.00+0.04+0.02}_{-0.00-0.02-0.01}$\\
RQM\cite{Faustov:2022ybm}&$-9.7\times10^{-7}$&$-0.013$&$-0.36$\\
PQCD\cite{Hu:2019bdf}&$-$&$-$&$0.36$\\
PQCD+Lattice\cite{Hu:2019bdf}&$-$&$-$&$0.36$\\
\hline				
				Channels&$B^{0}\to D^{*-}(1S)e^{+}\nu_{e}$&$B^{0}\to D^{*-}(1S)\mu^{+}\nu_{\mu}$&$B^{0}\to D^{*-}(1S)\tau^{+}\nu_{\tau}$\\
				\hline
				$\mathcal{A}_{FB}$&$-0.20^{+0.00+0.01+0.01}_{-0.00-0.00-0.01}$&$-0.20^{+0.00+0.01+0.01}_{-0.00-0.00-0.01}$&$-0.14^{+0.00+0.00+0.01}_{-0.00-0.00-0.01}$\\
				CQM\cite{Ivanov:2015tru}&$0.19$&$-$&$0.027$\\
				PQCD\cite{Hu:2019bdf}&$-$&$-$&$-0.085$\\
				PQCD+Lattice\cite{Hu:2019bdf}&$-$&$-$&$-0.054$\\
		    	HQET\cite{Huang:2018nnq}&$-$&$-$&$-0.084$\\
				\hline
				Channels&$B^{+}\to D^{*0}(1S)e^{+}\nu_{e}$&$B^{+}\to D^{*0}(1S)\mu^{+}\nu_{\mu}$&$B^{+}\to D^{*0}(1S)\tau^{+}\nu_{\tau}$\\
				\hline
				$\mathcal{A}_{FB}$&$-0.20^{+0.00+0.01+0.01}_{-0.00-0.00-0.01}$&$-0.20^{+0.00+0.01+0.01}_{-0.00-0.00-0.01}$&$-0.14^{+0.00+0.00+0.01}_{-0.00-0.00-0.01}$\\
				RQM\cite{Faustov:2022ybm}&$-0.22$&$-0.23$&$-0.32$\\
				\hline
				Channels&$B_{s}^{0}\to D_{s}^{*-}(1S)e^{+}\nu_{e}$&$B_{s}^{0}\to D_{s}^{*-}(1S)\mu^{+}\nu_{\mu}$&$B_{s}^{0}\to D_{s}^{*-}(1S)\tau^{+}\nu_{\tau}$\\
				\hline
				$\mathcal{A}_{FB}$&$-0.20^{+0.00+0.01+0.00}_{-0.00-0.00-0.01}$&$-0.20^{+0.00+0.01+0.02}_{-0.00-0.00-0.01}$&$-0.14^{+0.00+0.00+0.01}_{-0.00-0.00-0.01}$\\
				RQM\cite{Faustov:2022ybm}&$-0.26$&$-0.27$&$-0.32$\\
				PQCD\cite{Hu:2019bdf}&$-$&$-$&$-0.083$\\
				PQCD+Lattice\cite{Hu:2019bdf}&$-$&$-$&$-0.050$\\
				\hline\hline
			\end{tabular}\label{AFB1}}
	\end{center}
\end{table}

\begin{table}[H]
	\caption{Forward-backward asymmetries $\mathcal{A}_{\text{FB}}$ for the decays $B_{(s)} \to D_{(s)}(2S)\ell\nu_{\ell}$ and $B_{(s)} \to D^{*}_{(s)}(2S)\ell\nu_{\ell}$.}
	\begin{center}
		\scalebox{0.8}{
			\begin{tabular}{|c|c|c|c|}
				\hline\hline
				Channels  &$B^{0}\to D^{-}(2S) e^{+}{\nu}_{e}$&$B^{0}\to D^{-}(2S) \mu^{+}{\nu}_{\mu}$&$B^{0}\to D^{-}(2S) \tau^{+}{\nu}_{\tau}$\\
				\hline
				$\mathcal{A}_{\text{FB}}$&$6.60^{+0.02+0.08+0.08}_{-0.02-0.09-0.07}\times10^{-7}$&$0.02^{+0.00+0.00+0.00}_{-0.00-0.00-0.00}$&$0.35^{+0.00+0.05+0.05}_{-0.00-0.06-0.06}$\\
				\hline
				Channels&$B^{+}\to D^{0}(2S)e^{+}\nu_{e}$&$B^{+}\to D^{0}(2S)\mu^{+}\nu_{\mu}$&$B^{+}\to D^{0}(2S)\tau^{+}\nu_{\tau}$\\
				\hline
				$\mathcal{A}_{\text{FB}}$&$6.60^{+0.02+0.09+0.07}_{-0.02-0.08-0.08}\times10^{-7}$&$0.02^{+0.00+0.00+0.00}_{-0.00-0.00-0.00}$&$0.35^{+0.00+0.05+0.05}_{-0.00-0.06-0.06}$\\
			    \hline
				Channels&$B_{s}^{0}\to D_{s}^{-}(2S)e^{+}\nu_{e}$&$B_{s}^{0}\to D_{s}^{-}(2S)\mu^{+}\nu_{\mu}$&$B_{s}^{0}\to D_{s}^{-}(2S)\tau^{+}\nu_{\tau}$\\
				\hline
				$\mathcal{A}_{\text{FB}}$&$6.63^{+0.02+1.09+0.58}_{-0.02-1.04-1.03}\times10^{-7}$&$0.02^{+0.00+0.00+0.00}_{-0.00-0.00-0.00}$&$0.37^{+0.00+0.07+0.06}_{-0.00-0.08-0.08}$\\
				\hline
				Channels&$B^{0}\to D^{*-}(2S)e^{+}\nu_{e}$&$B^{0}\to D^{*-}(2S)\mu^{+}\nu_{\mu}$&$B^{0}\to D^{*-}(2S)\tau^{+}\nu_{\tau}$\\
				\hline
				$\mathcal{A}_{\text{FB}}$&$-0.12^{+0.00+0.01+0.06}_{-0.00-0.02-0.05}$&$-0.11^{+0.00+0.01+0.06}_{-0.00-0.02-0.05}$&$-0.05^{+0.00+0.00+0.03}_{-0.00-0.02-0.03}$\\
				\hline
				Channels&$B^{+}\to D^{*0}(2S)e^{+}\nu_{e}$&$B^{+}\to D^{*0}(2S)\mu^{+}\nu_{\mu}$&$B^{+}\to D^{*0}(2S)\tau^{+}\nu_{\tau}$\\
				\hline
				$\mathcal{A}_{\text{FB}}$&$-0.12^{+0.00+0.01+0.06}_{-0.00-0.02-0.05}$&$-0.12^{+0.00+0.01+0.06}_{-0.00-0.02-0.05}$&$-0.05^{+0.00+0.00+0.03}_{-0.00-0.02-0.03}$\\
				\hline
				Channels&$B_{s}^{0}\to D_{s}^{*-}(2S)e^{+}\nu_{e}$&$B_{s}^{0}\to D_{s}^{*-}(2S)\mu^{+}\nu_{\mu}$&$B_{s}^{0}\to D_{s}^{*-}(2S)\tau^{+}\nu_{\tau}$\\
				\hline
				$\mathcal{A}_{\text{FB}}$&$-0.13^{+0.00+0.04+0.06}_{-0.00-0.06-0.05}$&$-0.13^{+0.00+0.04+0.06}_{-0.00-0.06-0.05}$&$-0.07^{+0.00+0.02+0.04}_{-0.00-0.06-0.03}$\\
				\hline\hline
			\end{tabular}\label{AFB2}}
	\end{center}
\end{table}

\section{Summary}\label{sum}
Inspired by the recent experimental anomalies in B meson decays, especially the improved precision of LFU measurements, we
investigate the semileptonic decays $B_{(s)} \to D_{(s)}(nS)\ell \nu_{\ell}$ and $B_{(s)} \to  D^{*}_{(s)}(nS)\ell \nu_{\ell}$ with $n=1,2$ using the CLFQM, and calculate several physical observables, including the branching ratio, $\mathcal{R}(D^{(*)})$, the forward-backward asymmetry $\mathcal{A}_{\text{FB}}$ and longitudinal polarization fraction $f_L$. Our predictions for the branching ratios of the decays $B_{(s)}\to D_{(s)}(1S)\ell^\prime\nu_{\ell^\prime}, D^*_{(s)}(1S)\ell^\prime\nu_{\ell^\prime}$ agree well with data, and the deviations
are no more than $1.5\sigma$, reaching as low as $0.3\sigma$ for the decays $B_{(s)}\to D^*_{(s)}(1S)\ell^\prime\nu_{\ell^\prime}$. While the branching ratios of the decays $B_{(s)}\to D_{(s)}(1S)\tau\nu_\tau, D^{*}_{(s)}(1S)\tau\nu_\tau$ are systematically smaller than the experimental data. The deviations between our predictions and the data for the decays $B^0\to D^{(*)-}\tau^+\nu_\tau, B^+\to D^{*0}\tau^+\nu_\tau$ exceed $2\sigma$, or even $3\sigma$. Accordingly, the extracted values of $\mathcal{R}(D)$ and $\mathcal{R}(D^*)$ are naturally lower than the experimental measurements. Nevertheless, such tensions display a gradual downward trend. The deviations between our results and the latest LHCb data for $\mathcal{R}(D)$ and $\mathcal{R}(D^*)$ are $0.16\sigma$ and $1.5\sigma$, respectively. In contrast, the tensions compared with the current HFLAV experimental world averages still amount to $3.1\sigma$ and $2.1\sigma$, respectively. Meanwhile, the branching ratios of the decays $B\to  D^{(*)}(2S)\ell\nu_\ell$ are also calculated, which lie in the range $10^{-4}$ to $10^{-3}$ and are about one order of magnitude larger than those obtained from the BS equation. For decays $B_s\to  D^{(*)}_s(2S)\ell\nu_\ell$, our predictions are $3$ to $5$ times those given by the BS equation, and are consistent with the RQM calculations. Furthermore, we present the forward-backward asymmetries $\mathcal{A}_{FB}$ and the longitudinal polarization fractions $f_{L}$ for relevant decays, which can be tested in future LHCb and Belle II experiments. 
\begin{table}[H]
	\caption{The longitudinal polarization fractions $f_{L}$ for the decays $B_{(s)} \to D^{*}_{(s)}(1S,2S)\ell\nu_{\ell}$ in Region 1,  Region 2 and total physical region, respectively.}
	\begin{center}
		\scalebox{0.8}{
			\begin{tabular}{|c|c|c|c|c|c|c|c|}
				\hline\hline
				Observables&Region 1&Region 2&Total&Observables&Region 1&Region 2&Total\\
				\hline\hline
				$f_{L}(B^{0}\to D^{*-}(1S)l^{'+}\nu_{l^{'}})$&$0.35$&$0.20$&$0.55^{+0.00+0.00+0.03}_{-0.00-0.04-0.02}$&$f_{L}(B^{0}\to D^{*-}(2S)l^{'+}\nu_{l^{'}})$&$0.50$&$0.18$&$0.68^{+0.00+0.06+0.03}_{-0.00-0.05-0.03}$\\
				\hline
				$f_{L}(B^{0}\to D^{*-}(1S)\tau^{+}\nu_{\tau})$&$0.20$&$0.25$&$0.45^{+0.00+0.00+0.01}_{-0.00-0.03-0.02}$&$f_{L}(B^{0}\to D^{*-}(2S)\tau^{+}\nu_{\tau})$&$0.22$&$0.29$&$0.51^{+0.00+0.03+0.01}_{-0.00-0.06-0.02}$\\
				\hline
				PQCD\cite{Hu:2019bdf}&$-$&$-$&$0.42$&$-$&$-$&$-$&$-$\\
				PQCD+Lattice\cite{Hu:2019bdf}&$-$&$-$&$0.43$&$-$&$-$&$-$&$-$\\
				MIA\cite{Bhattacharya:2018kig}&$-$&$-$&$0.457$&$-$&$-$&$-$&$-$\\
				HQET\cite{Huang:2018nnq}&$-$&$-$&$0.441$&$-$&$-$&$-$&$-$\\
				Belle\cite{Belle:2019ewo}&$-$&$-$&$0.60\pm0.08\pm0.04$&$-$&$-$&$-$&$-$\\
				LHCb\cite{LHCb:2023ssl}&$-$&$-$&$0.41\pm0.06\pm0.03$&$-$&$-$&$-$&$-$\\
				\hline\hline
				Observables&Region 1&Region 2&Total&Observables&Region 1&Region 2&Total\\
				\hline\hline
				$f_{L}(B^{+}\to D^{*0}(1S)l^{'+}\nu_{l^{'}})$&$0.35$&$0.20$&$0.55^{+0.00+0.00+0.03}_{-0.00-0.03-0.02}$&$f_{L}(B^{+}\to D^{*0}(2S)l^{'+}\nu_{l^{'}})$&$0.49$&$0.18$&$0.68^{+0.00+0.05+0.03}_{-0.00-0.06-0.03}$\\
				\hline
				RQM\cite{Faustov:2022ybm}&$-$&$-$&$0.55$&$-$&$-$&$-$&$-$\\
				\hline
				$f_{L}(B^{+}\to D^{*0}(1S)\tau^{+}\nu_{\tau})$&$0.20$&$0.25$&$0.45^{+0.00+0.00+0.01}_{-0.00-0.03-0.06}$&$f_{L}(B^{+}\to D^{*0}(2S)\tau^{+}\nu_{\tau})$&$0.22$&$0.29$&$0.51^{+0.00+0.4+0.03}_{-0.00-0.06-0.03}$\\
				\hline
				RQM\cite{Faustov:2022ybm}&$-$&$-$&$0.47$&$-$&$-$&$-$&$-$\\
				\hline\hline
				Observables&Region 1&Region 2&Total&Observables&Region 1&Region 2&Total\\
				\hline\hline
				$f_{L}(B_{s}^{0}\to D_{s}^{*-}(1S)l^{'+}\nu_{l^{'}})$&$0.34$&$0.20$&$0.54^{+0.00+0.04+0.05}_{-0.00-0.03-0.07}$&$f_{L}(B_{s}^{0}\to D_{s}^{*-}(2S)l^{'+}\nu_{l^{'}})$&$0.51$&$0.17$&$0.68^{+0.00+0.31+0.15}_{-0.00-0.20-0.22}$\\
				\hline
				RQM\cite{Faustov:2022ybm} &$-$&$-$&$0.49$&$-$&$-$&$-$&$-$\\
				\hline
				$f_{L}(B_{s}^{0}\to D_{s}^{*-}(1S)\tau^{+}\nu_{\tau})$&$0.20$&$0.25$&$0.45^{+0.00+0.00+0.05}_{-0.00-0.02-0.03}$&$f_{L}(B_{s}^{0}\to D_{s}^{*-}(2S)\tau^{+}\nu_{\tau})$&$0.23$&$0.28$&$0.51^{+0.01+0.12+0.11}_{-0.01-0.18-0.11}$\\
				\hline
				PQCD\cite{Hu:2019bdf}&$-$&$-$&$0.42$&$-$&$-$&$-$&$-$\\
				PQCD+Lattice\cite{Hu:2019bdf}&$-$&$-$&$0.43$&$-$&$-$&$-$&$-$\\
				RQM\cite{Faustov:2022ybm}&$-$&$-$&$0.42$&$-$&$-$&$-$&$-$\\
				\hline\hline
			\end{tabular}\label{FLE}}
	\end{center}
\end{table}
%%%%%%%%%%%%%%%%%%%%%%%%%%%%%%%%%%%%%%%%%%%%%%%%%%%%%%%%%%%%%%%%%%%%%%%%%%%%%%%
\section*{Acknowledgment}
This work is partly supported by the National Natural Science Foundation of China under
Grant No. 11347030 and the Natural Science Foundation of Henan Province under grant
no. 232300420116, 242300421679, 252300421302.
\appendix
\section{$q^{2}$  dependence of the forward-backward asymmetries $\mathcal{A}_{FB}$}
\begin{figure}[H]
	\vspace{0.4cm}
	\centering
	\subfigure[]{\includegraphics[width=0.4\textwidth]{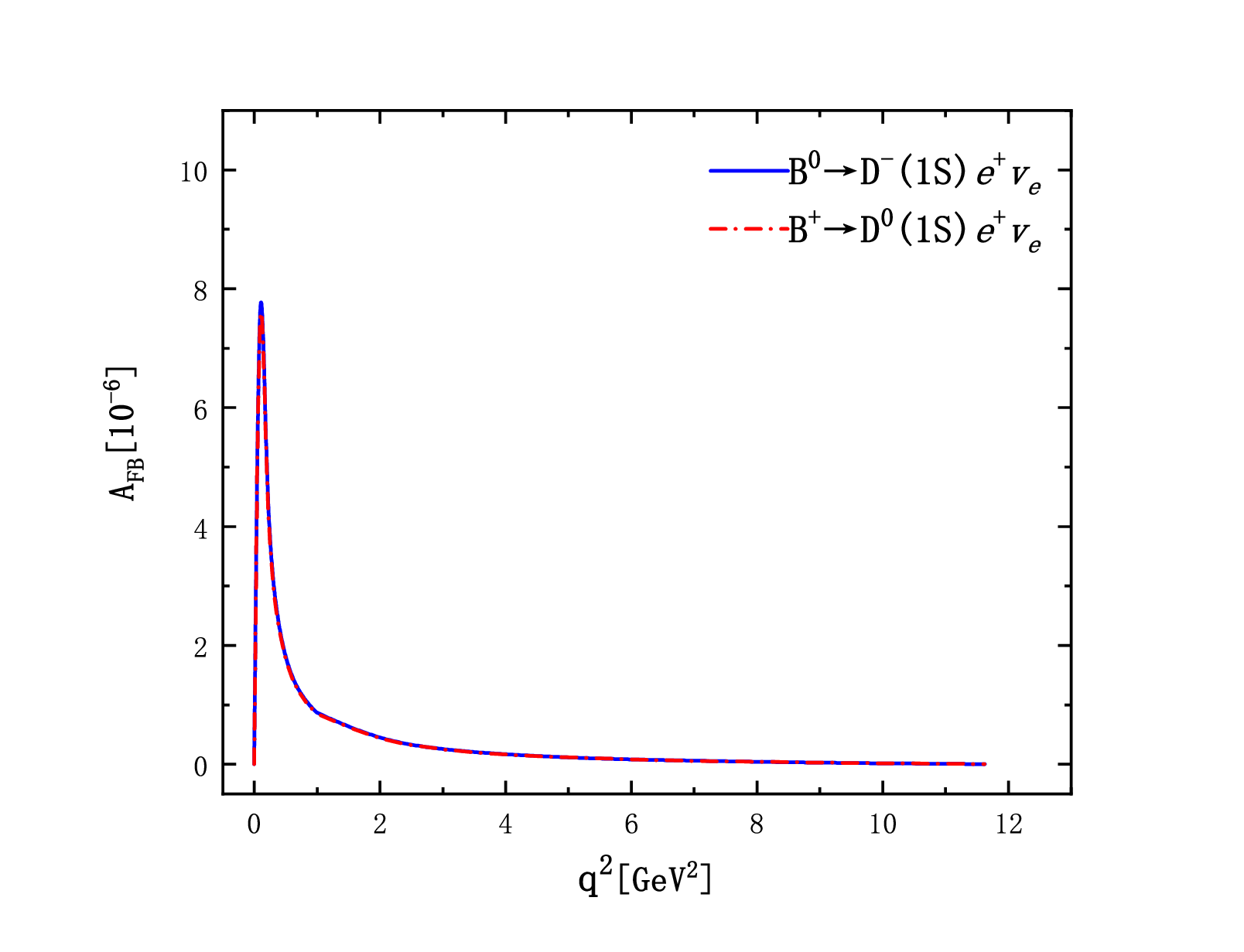}\quad}
	\subfigure[]{\includegraphics[width=0.4\textwidth]{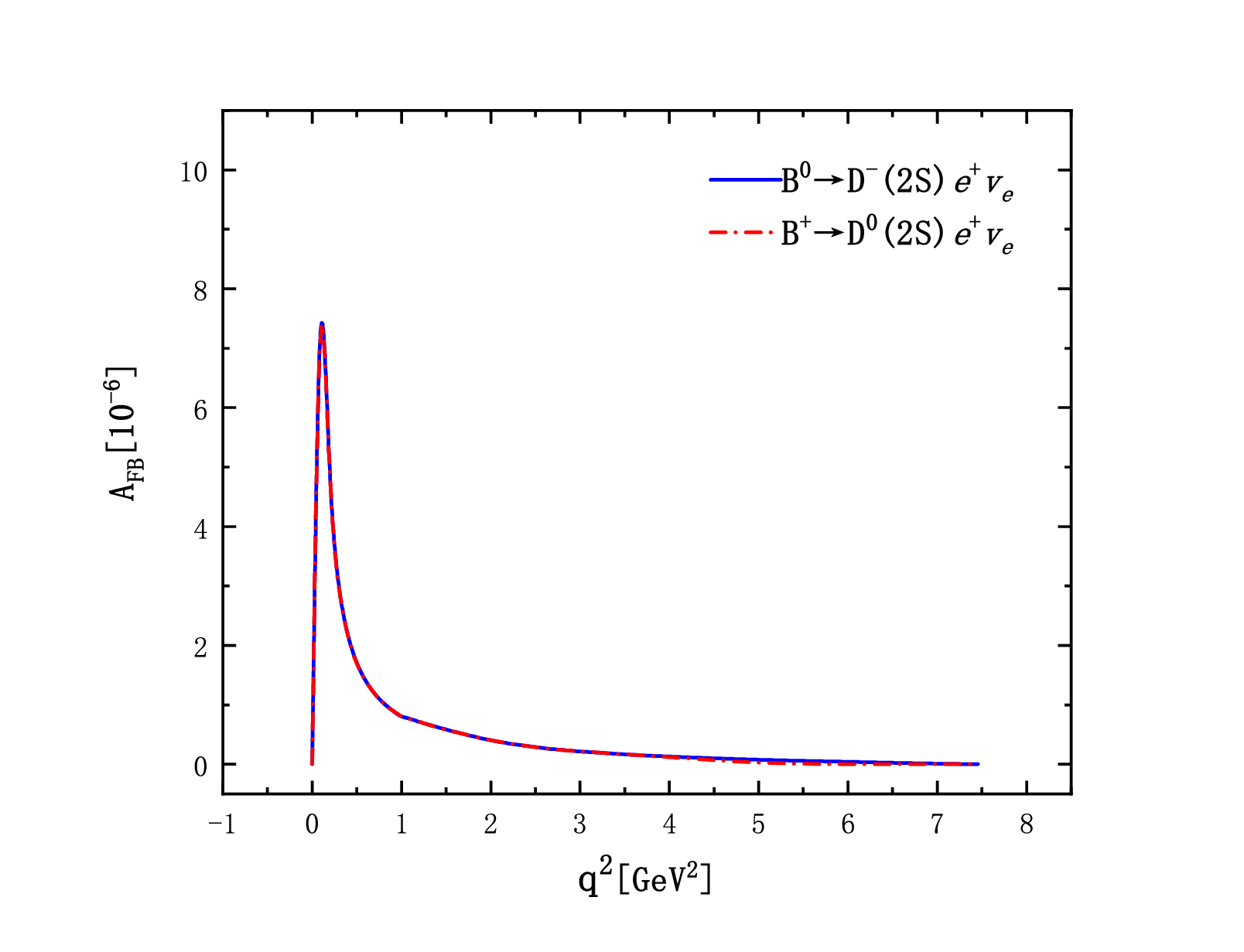}}\\
	\subfigure[]{\includegraphics[width=0.4\textwidth]{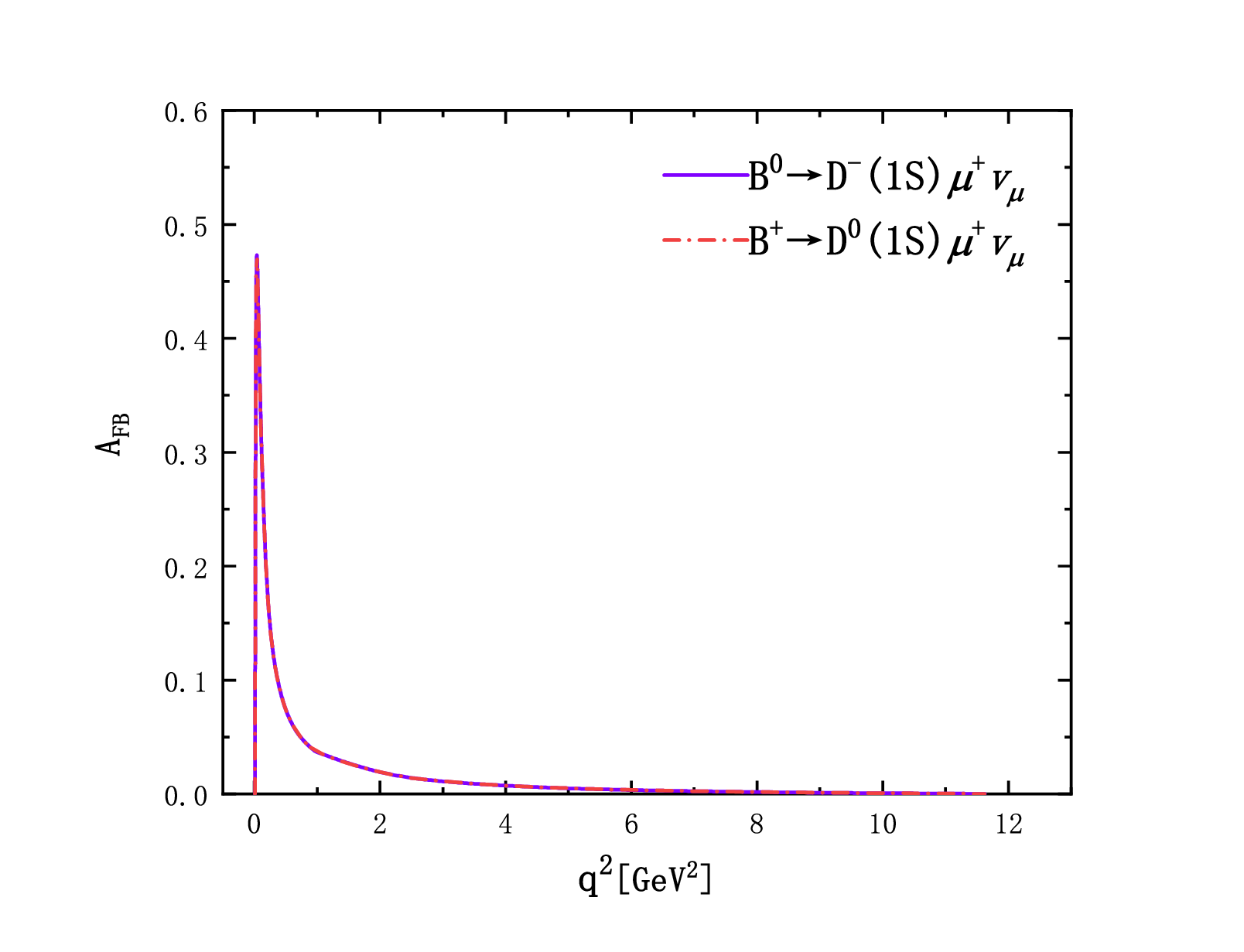}\quad}
	\subfigure[]{\includegraphics[width=0.4\textwidth]{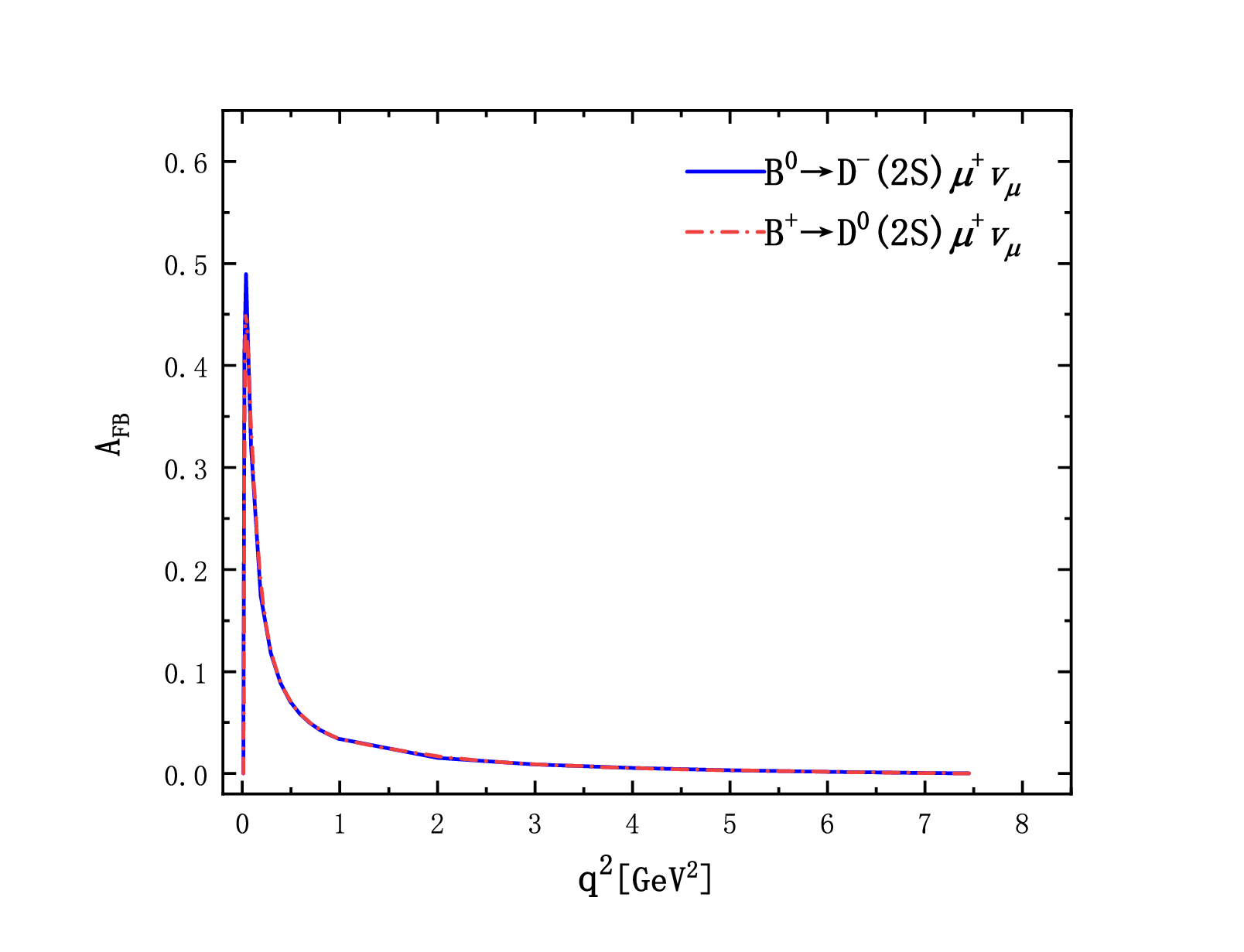}}\\
	\subfigure[]{\includegraphics[width=0.4\textwidth]{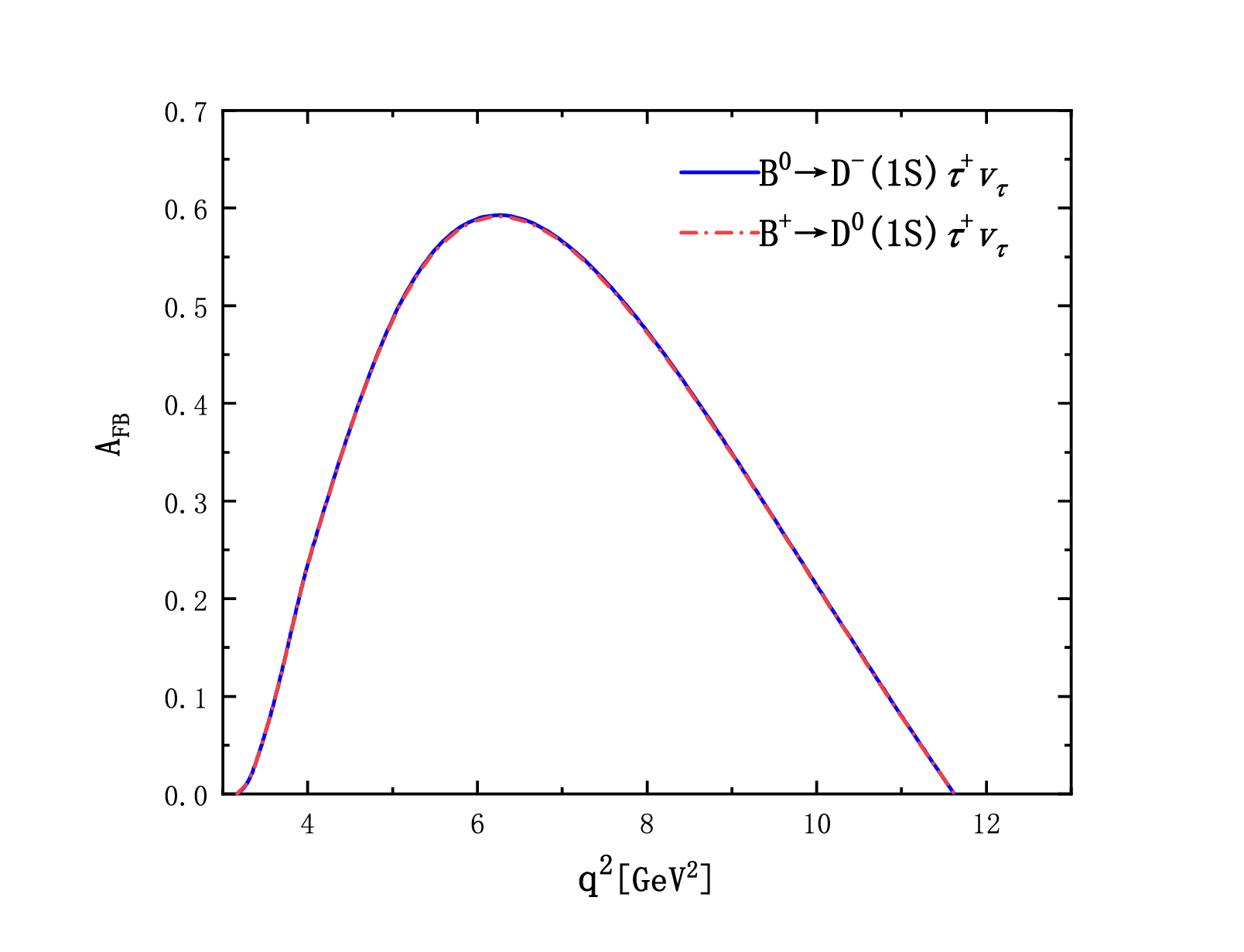}\quad}
	\subfigure[]{\includegraphics[width=0.4\textwidth]{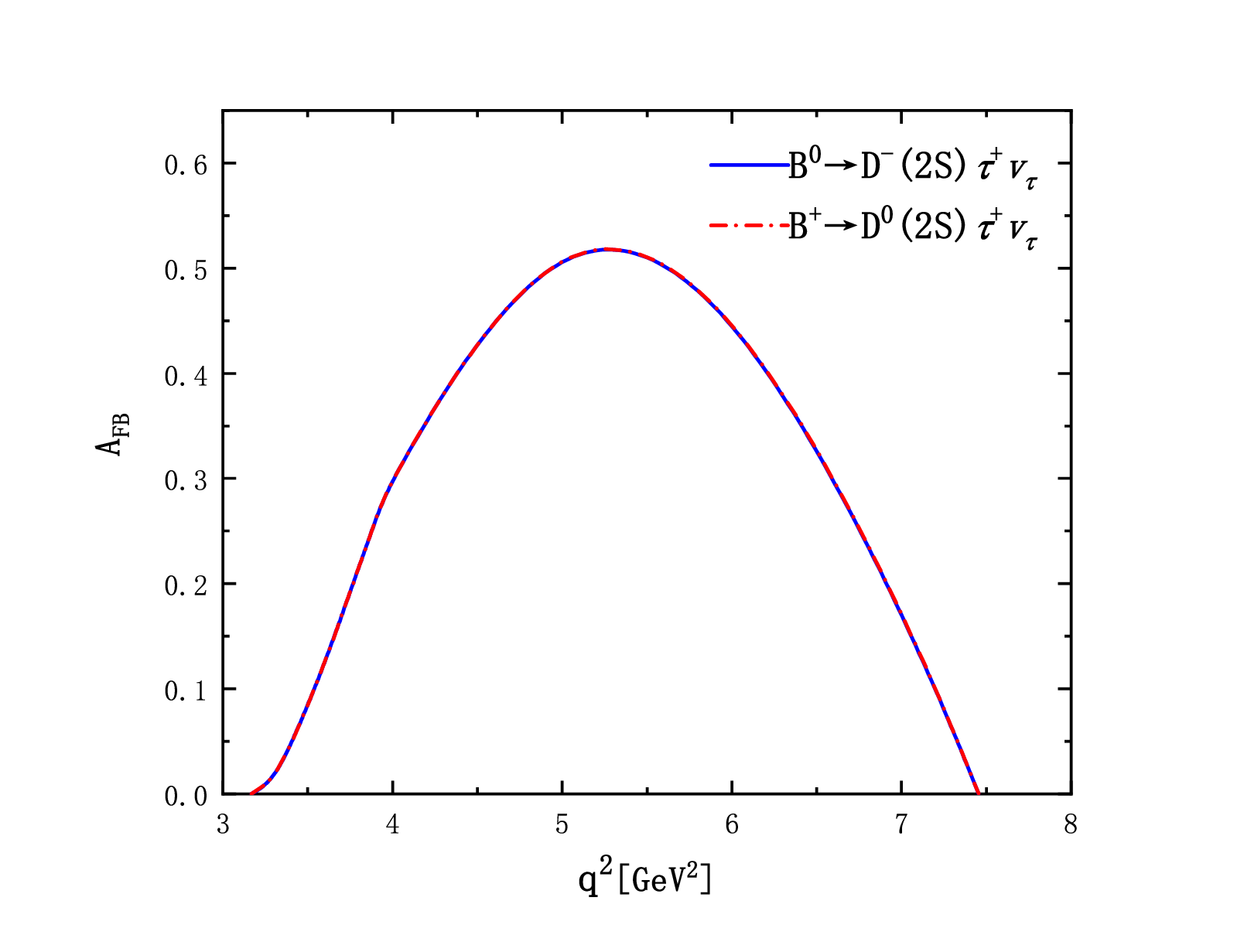}}\\
	\caption{The $q^{2}$  dependence of the forward-backward asymmetries $\mathcal{A}_{FB}$ for the decays $B\to D(1S)e^{+}\nu_{e}$ (a), $B\to D(2S)e^{+}\nu_{e}$ (b), $B\to D(1S)\mu^{+}\nu_{\mu}$ (c), $B\to D(2S)\mu^{+}\nu_{\mu}$ (d), $B\to D(1S)\tau^{+}\nu_{\tau}$ (e) and $B\to D(2S)\tau^{+}\nu_{\tau}$ (f).}\label{fig:T1}
\end{figure}

\begin{figure}[H]
	\vspace{0.4cm}
	\centering
	\subfigure[]{\includegraphics[width=0.4\textwidth]{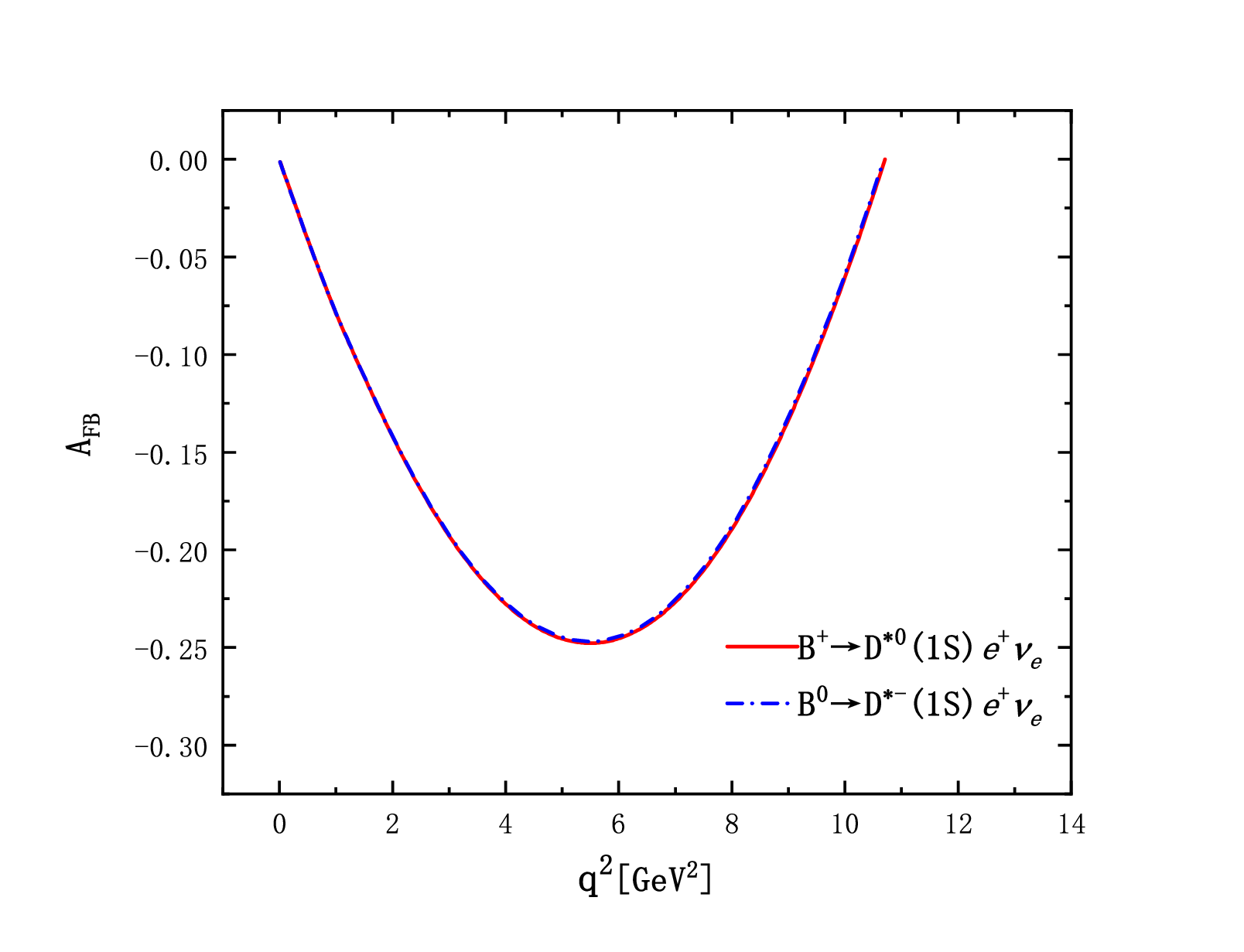}\quad}
    \subfigure[]{\includegraphics[width=0.4\textwidth]{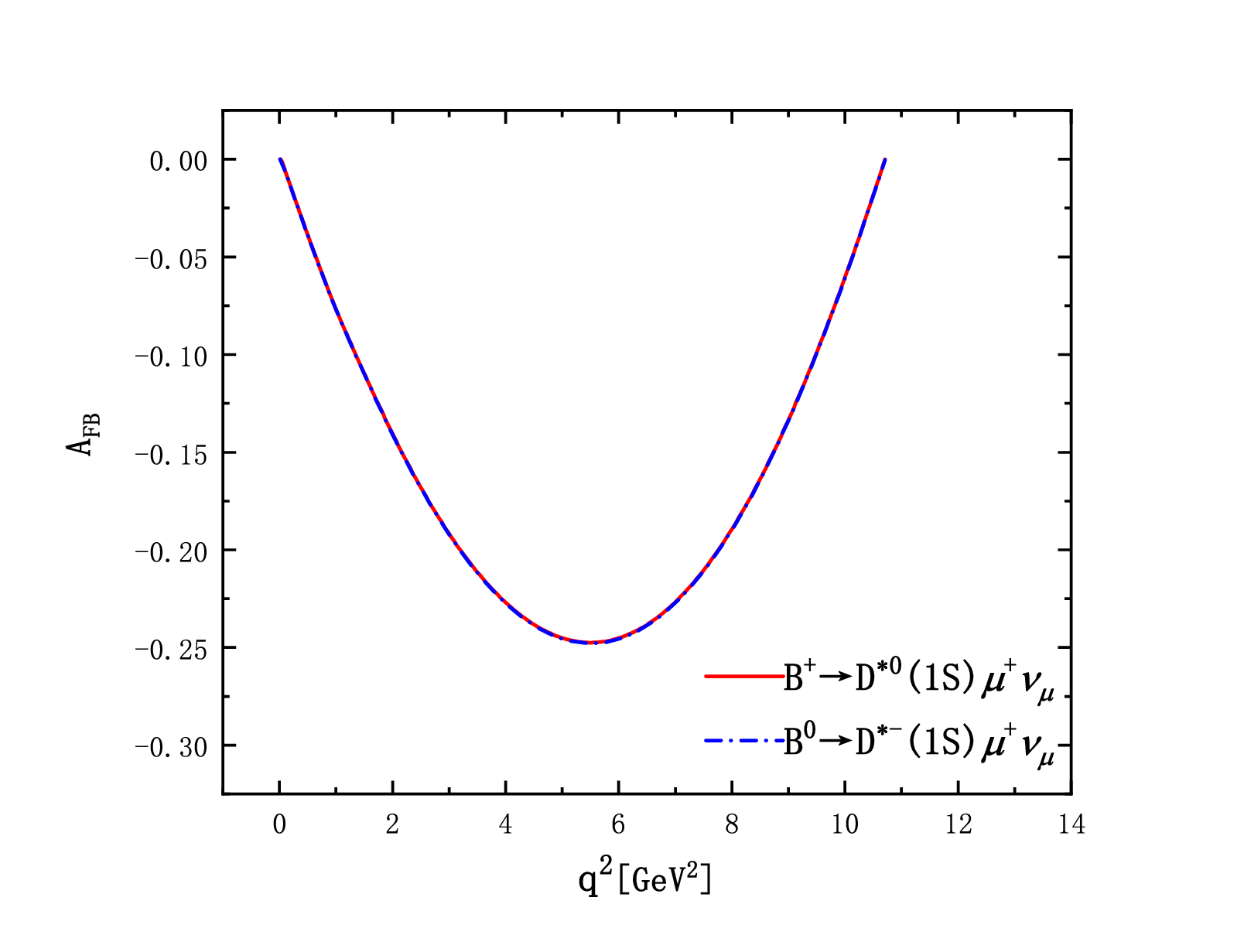}}\\
	\subfigure[]{\includegraphics[width=0.4\textwidth]{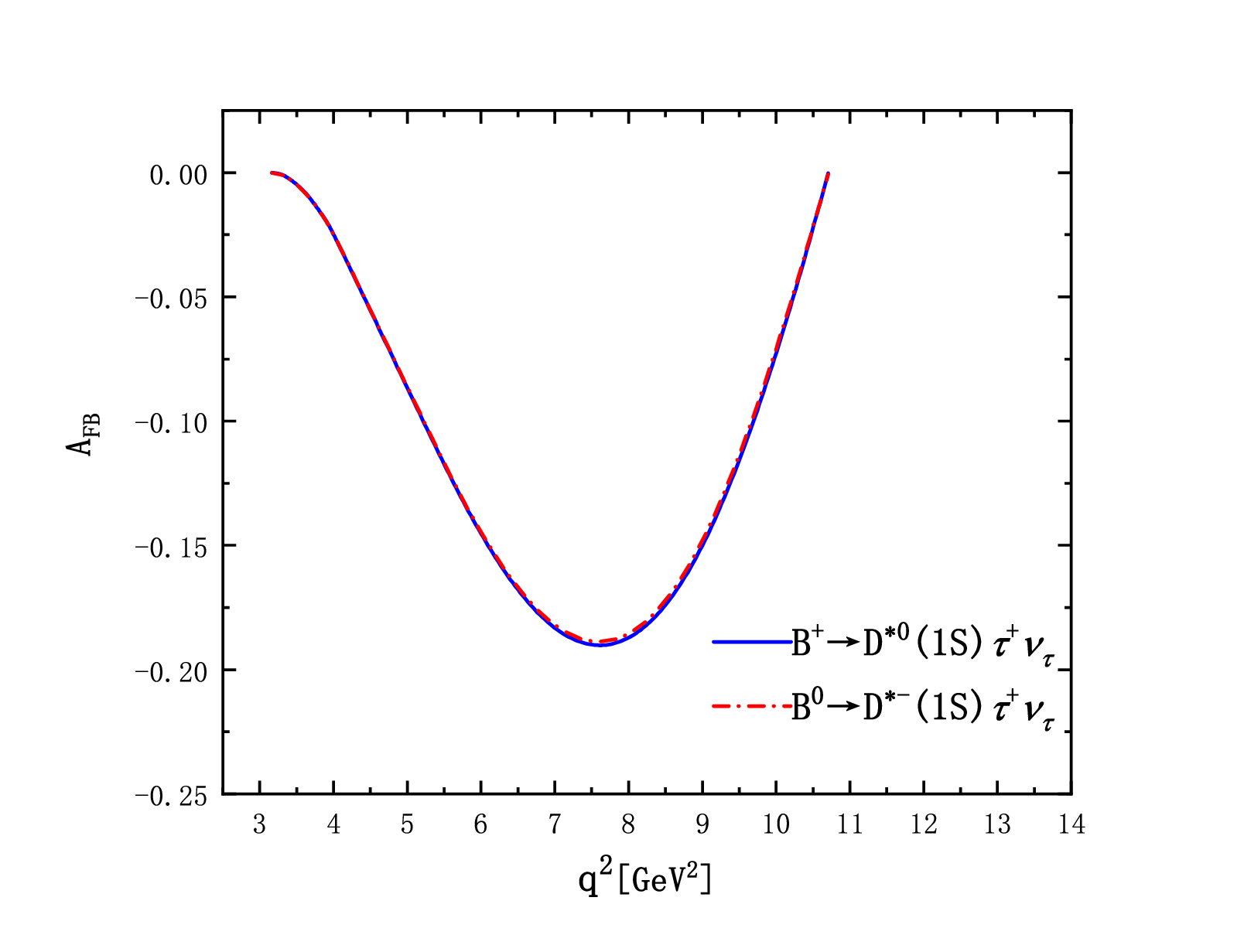}\quad}
	\subfigure[]{\includegraphics[width=0.4\textwidth]{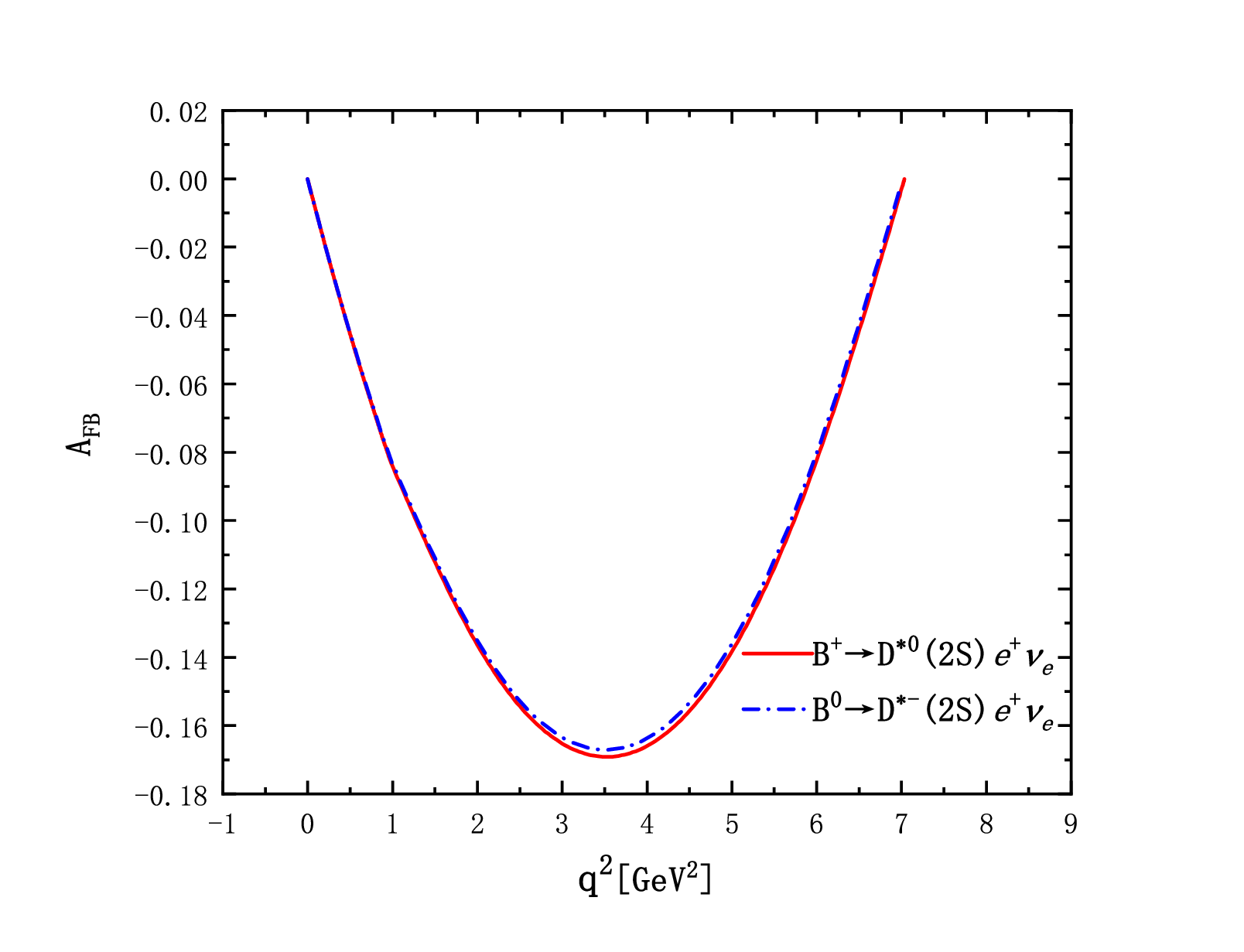}}\\
	\subfigure[]{\includegraphics[width=0.4\textwidth]{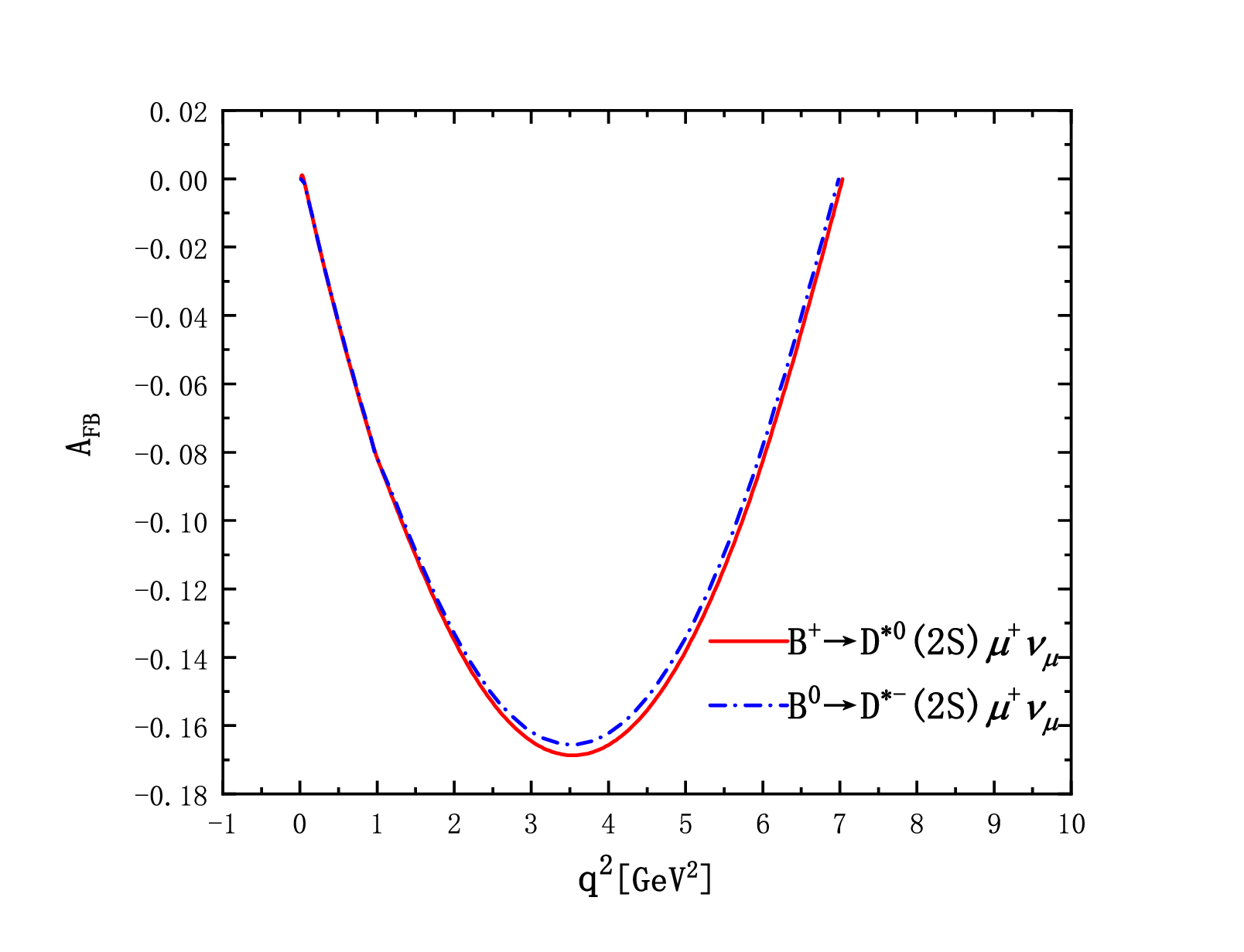}\quad}
	\subfigure[]{\includegraphics[width=0.4\textwidth]{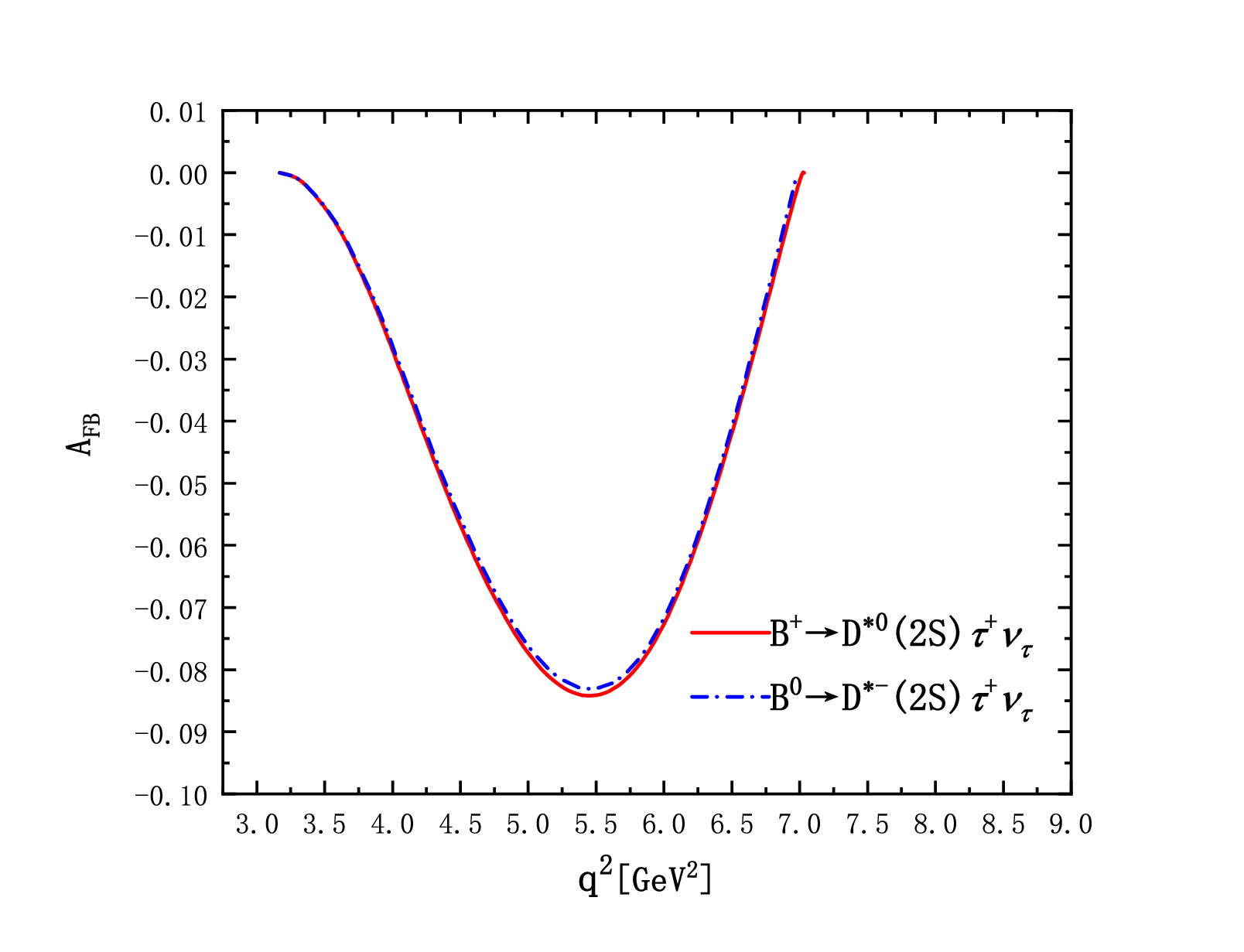}}\\
	\caption{The $q^{2}$  dependence of the forward-backward asymmetries $A_{FB}$ for the decays $B\to D^*(1S)e^{+}\nu_{e}$ (a), $B\to D^*(1S)\mu^{+}\nu_{\mu}$ (b), $B\to D^*(1S)\tau^{+}\nu_{\tau}$ (c), $B\to D^*(2S)e^{+}\nu_{e}$ (d), $B\to D^*(2S)\mu^{+}\nu_{\mu}$ (e) and $B\to D^*(2S)\tau^{+}\nu_{\tau}$ (f).}\label{fig:T3}
\end{figure}

\begin{figure}[H]
	\vspace{0.4cm}
	\centering
	\subfigure[]{\includegraphics[width=0.4\textwidth]{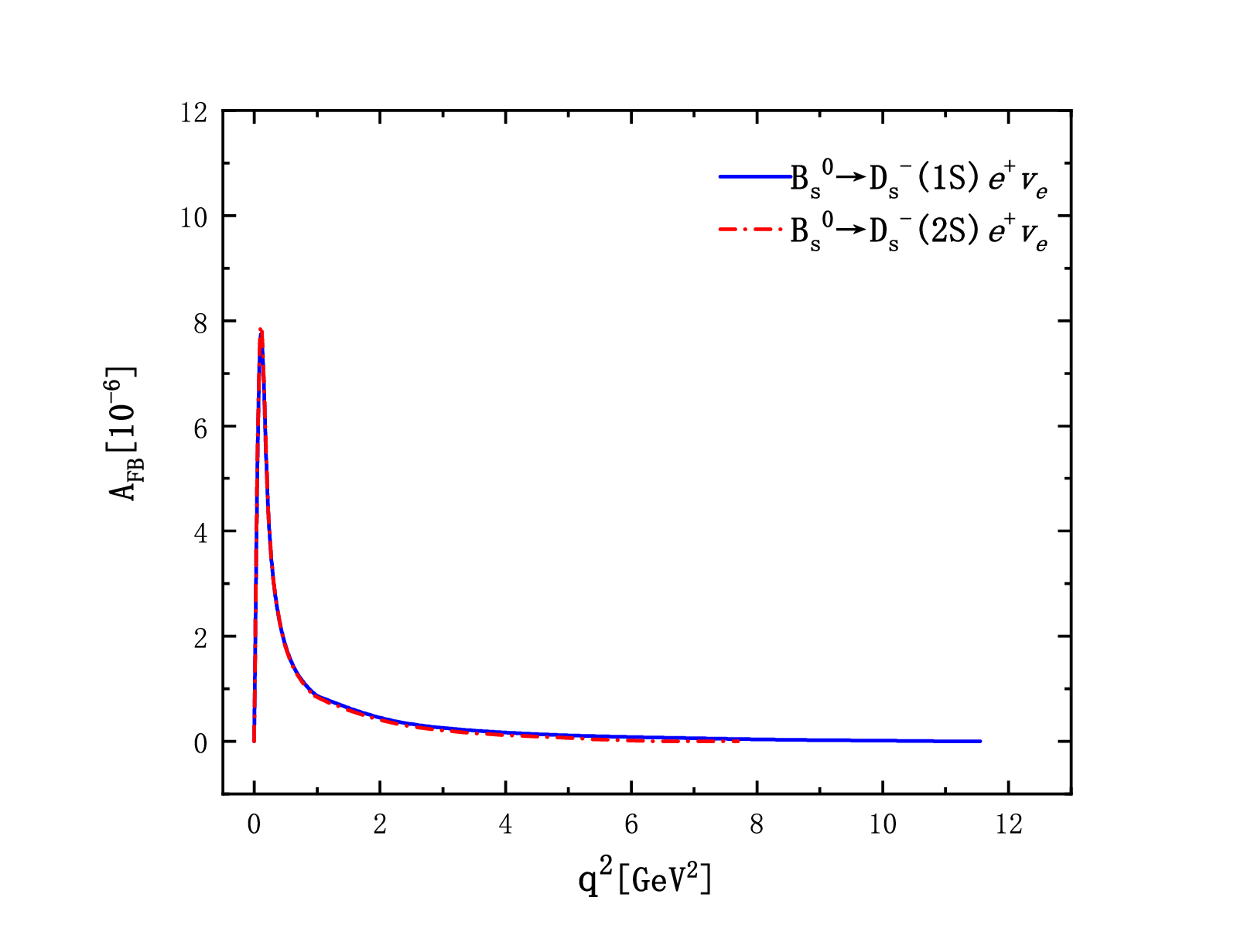}\quad}
	\subfigure[]{\includegraphics[width=0.4\textwidth]{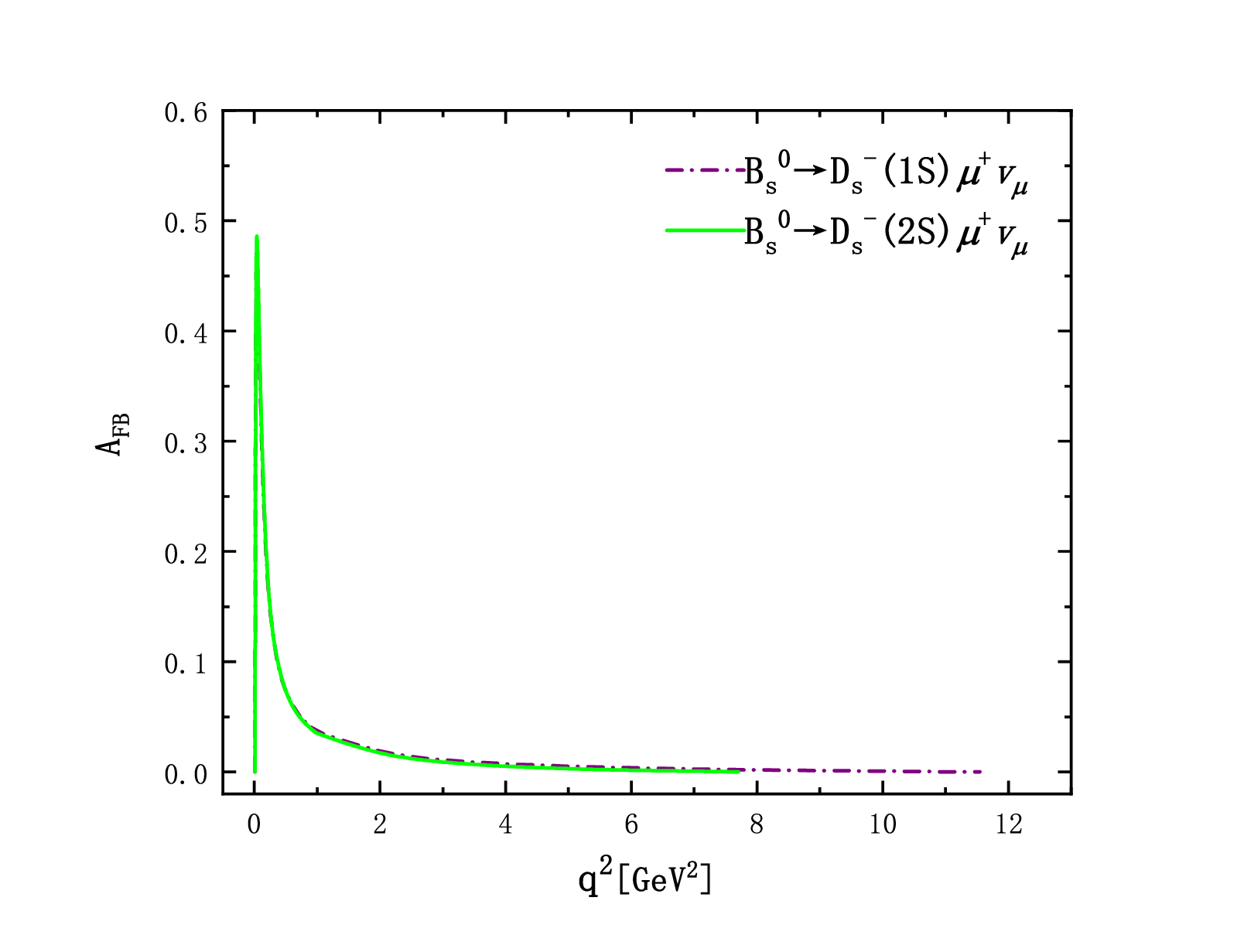}}\\
	\subfigure[]{\includegraphics[width=0.4\textwidth]{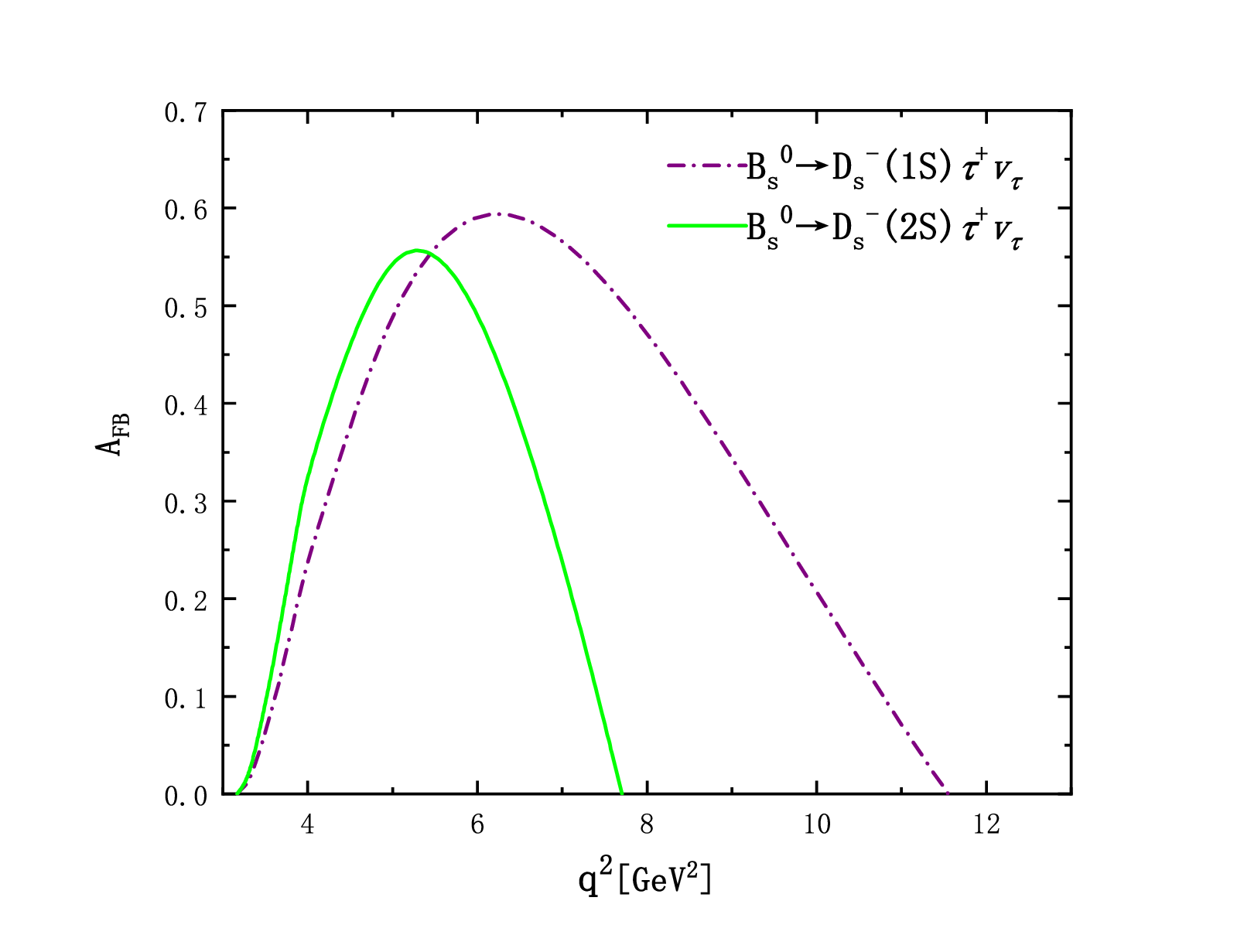}\quad}
	\subfigure[]{\includegraphics[width=0.4\textwidth]{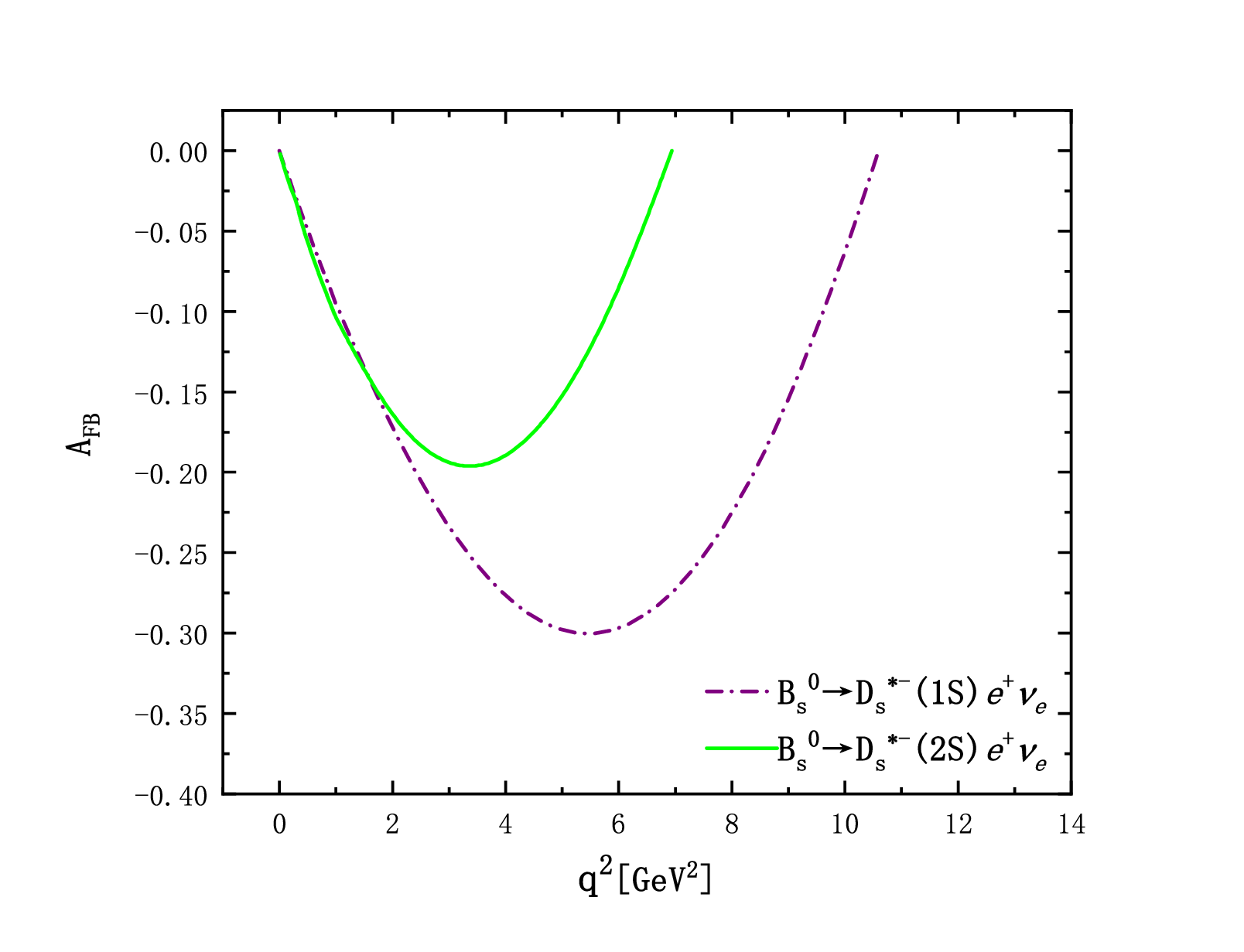}}\\
	\subfigure[]{\includegraphics[width=0.4\textwidth]{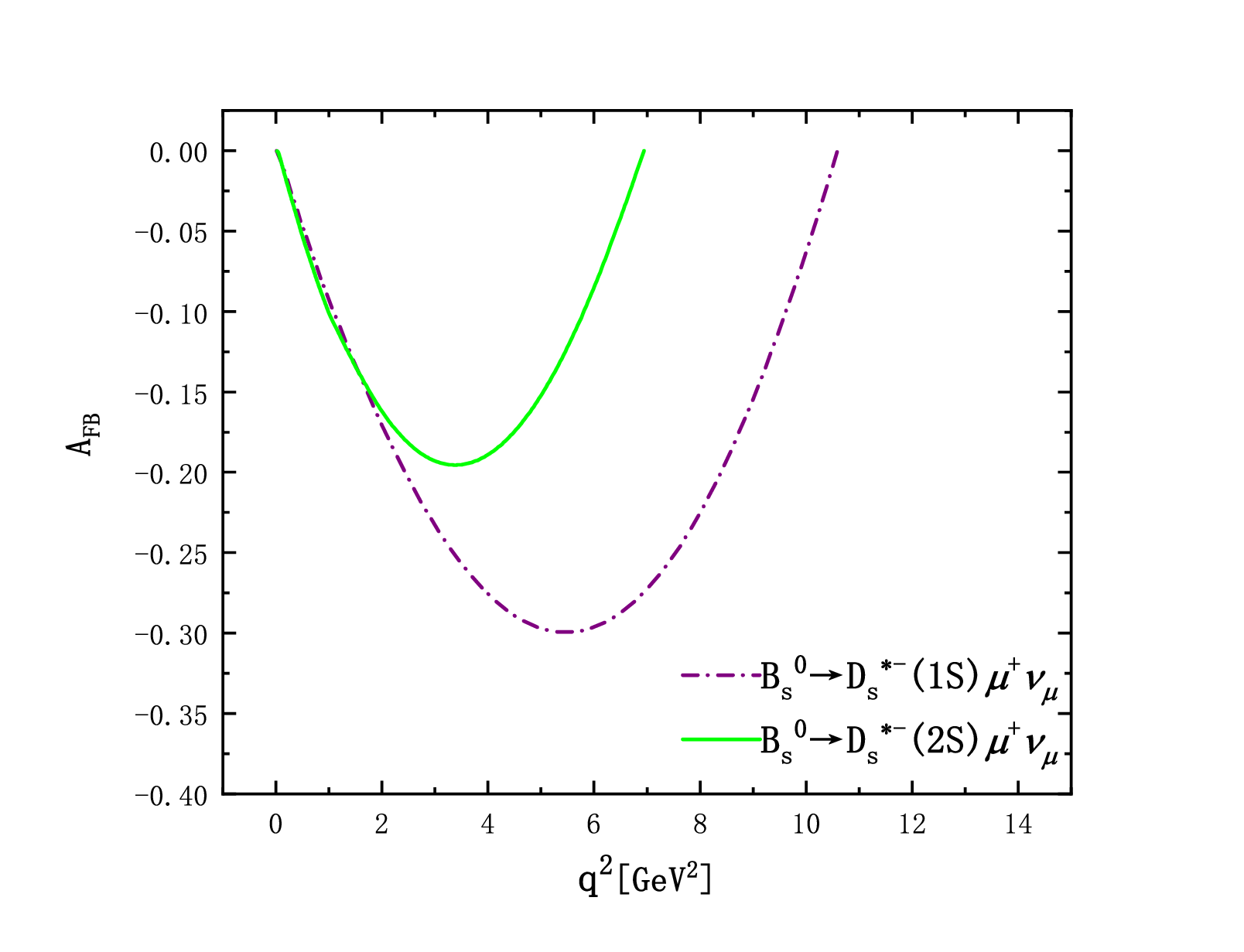}\quad}
	\subfigure[]{\includegraphics[width=0.4\textwidth]{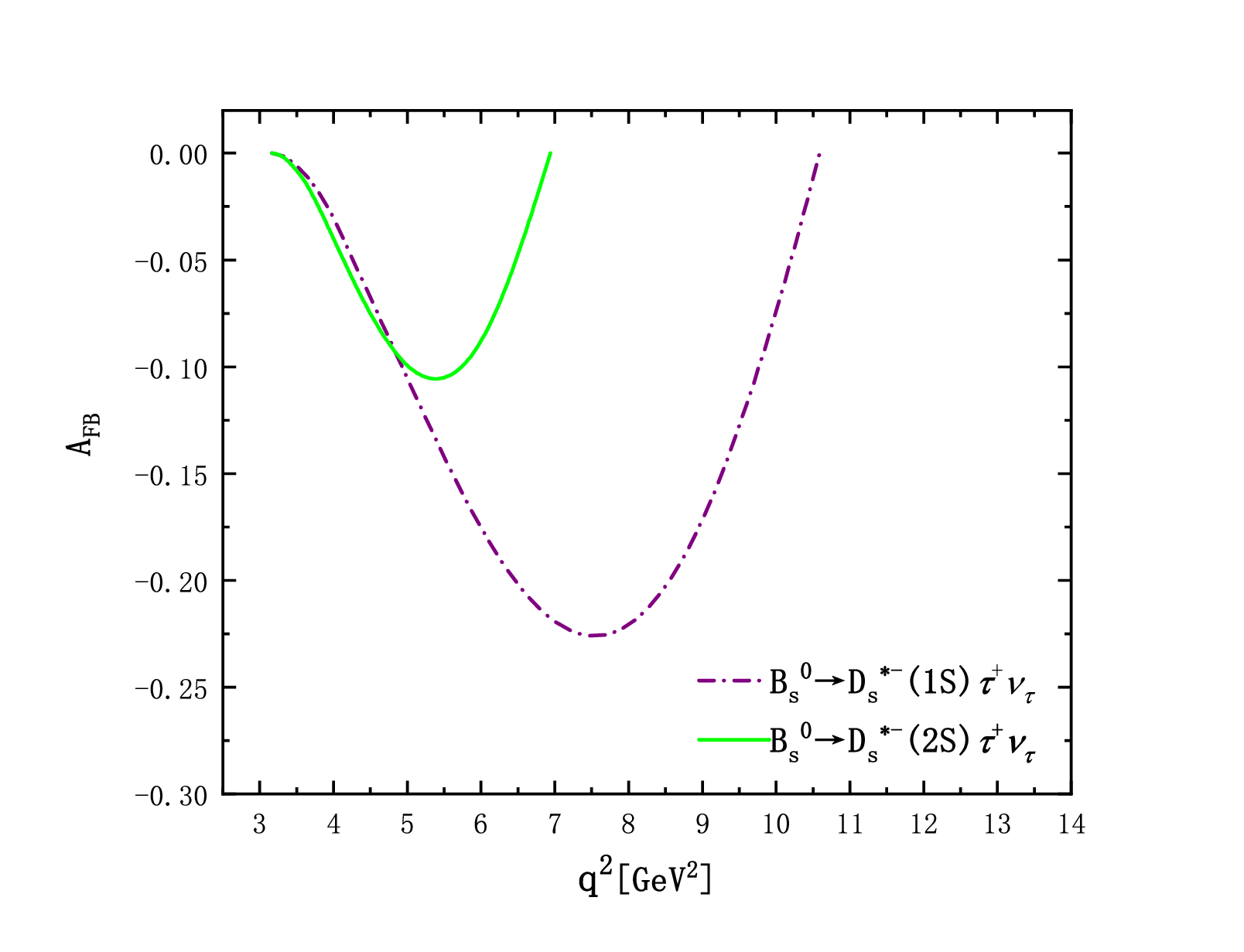}}\\
	\caption{The $q^{2}$ dependence of the forward-backward asymmetries $\mathcal{A}_{FB}$ for the decays $B_{s} \to D_{s}(1S,2S)e^{+}\nu_{e}$ (a), $B_{s} \to D_{s}(1S,2S)\mu^{+}\nu_{\mu}$ (b), $B_{s} \to D_{s}(1S,2S)\tau^{+}\nu_{\tau}$ (c), $B_{s}\to D^*_{s}(1S,2S)e^{+}\nu_{e}$ (d), $B_{s} \to D^*_{s}(1S,2S)\mu^{+}\nu_{\mu}$ (e) and $B_{s} \to D^*_{s}(1S,2S)\tau^{+}\nu_{\tau}$ (f).}\label{fig:T2}
\end{figure}

\begin{figure}[H]
	\vspace{0.4cm}
	\centering
	\subfigure[]{\includegraphics[width=0.4\textwidth]{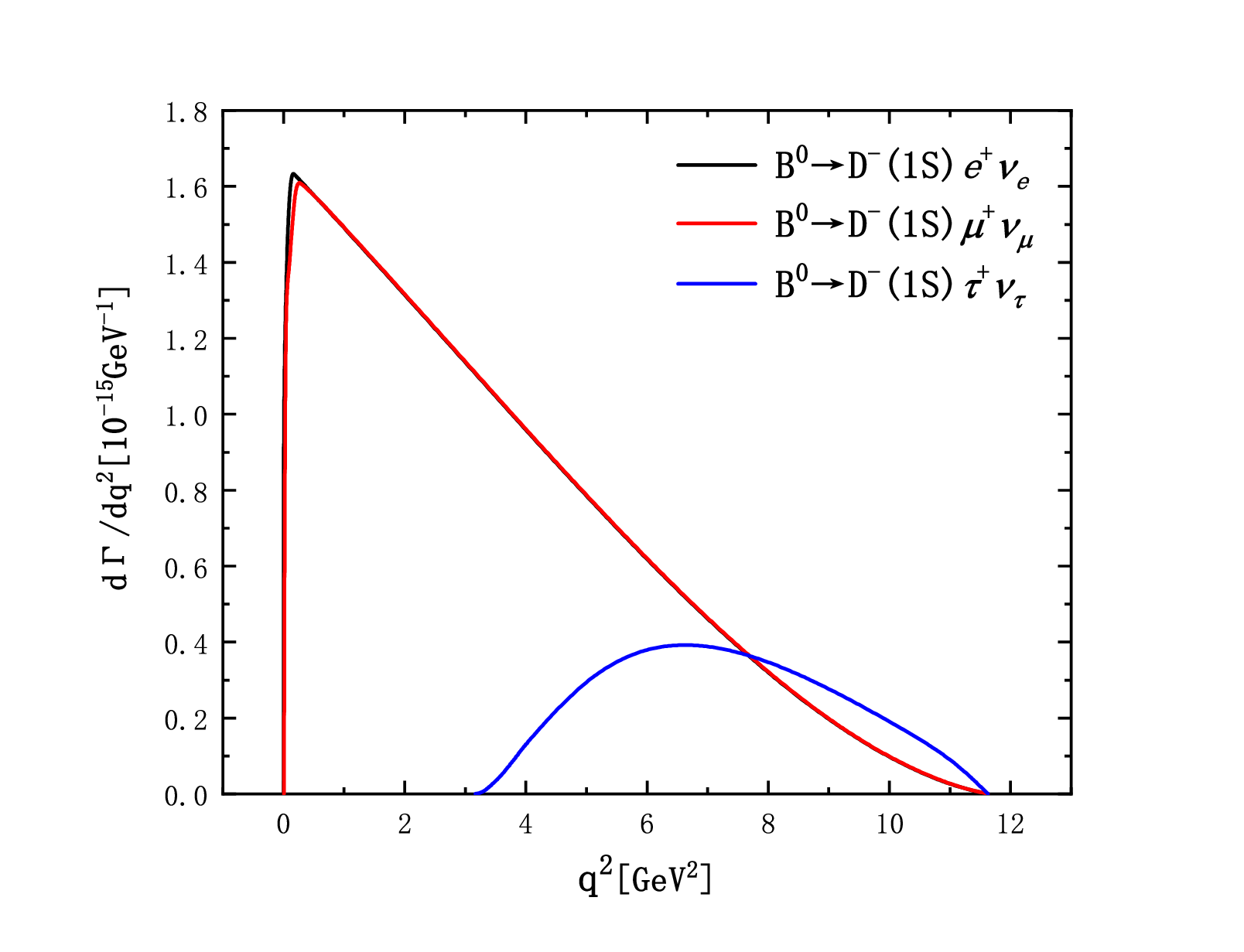}\quad}
	\subfigure[]{\includegraphics[width=0.4\textwidth]{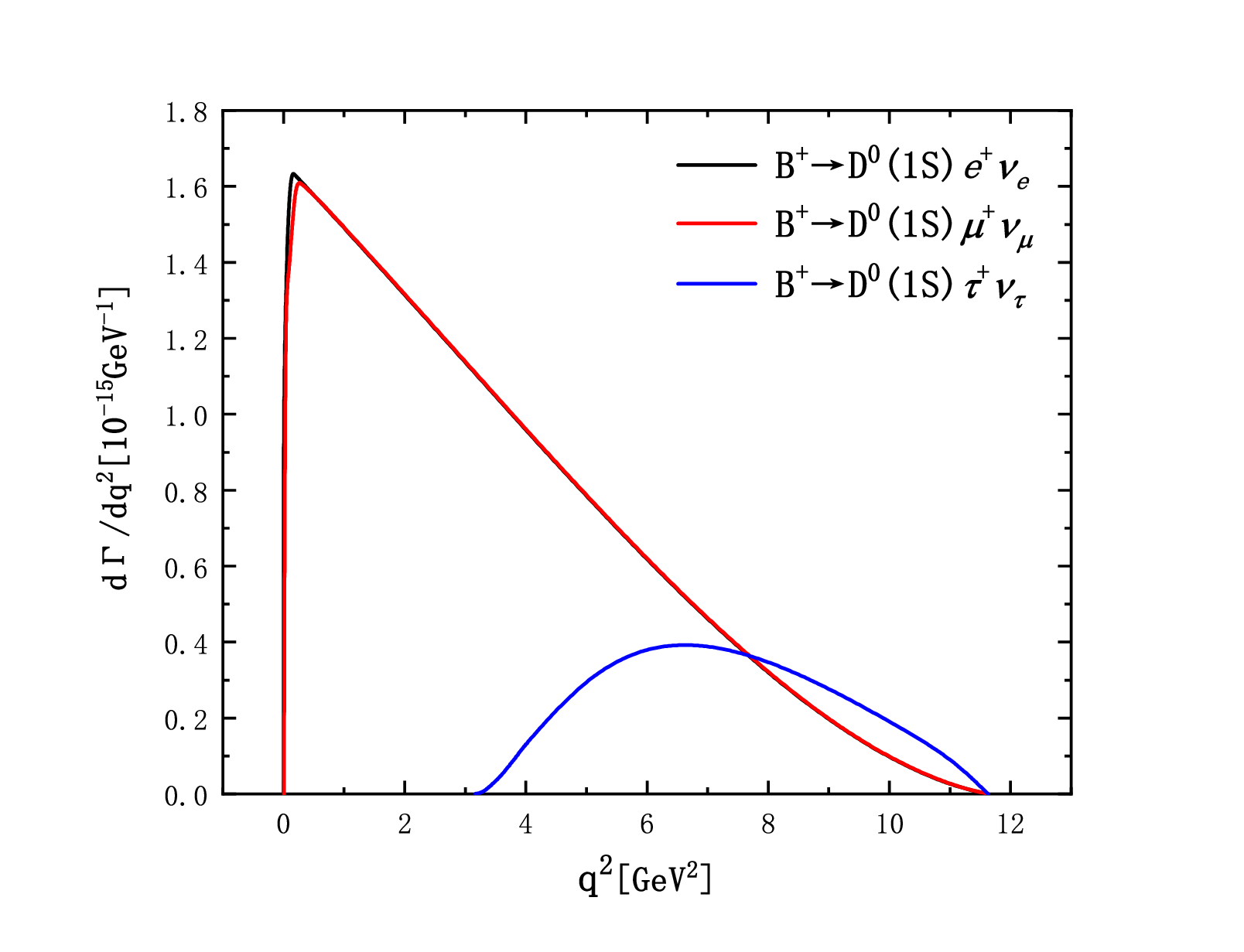}}\\
	\subfigure[]{\includegraphics[width=0.4\textwidth]{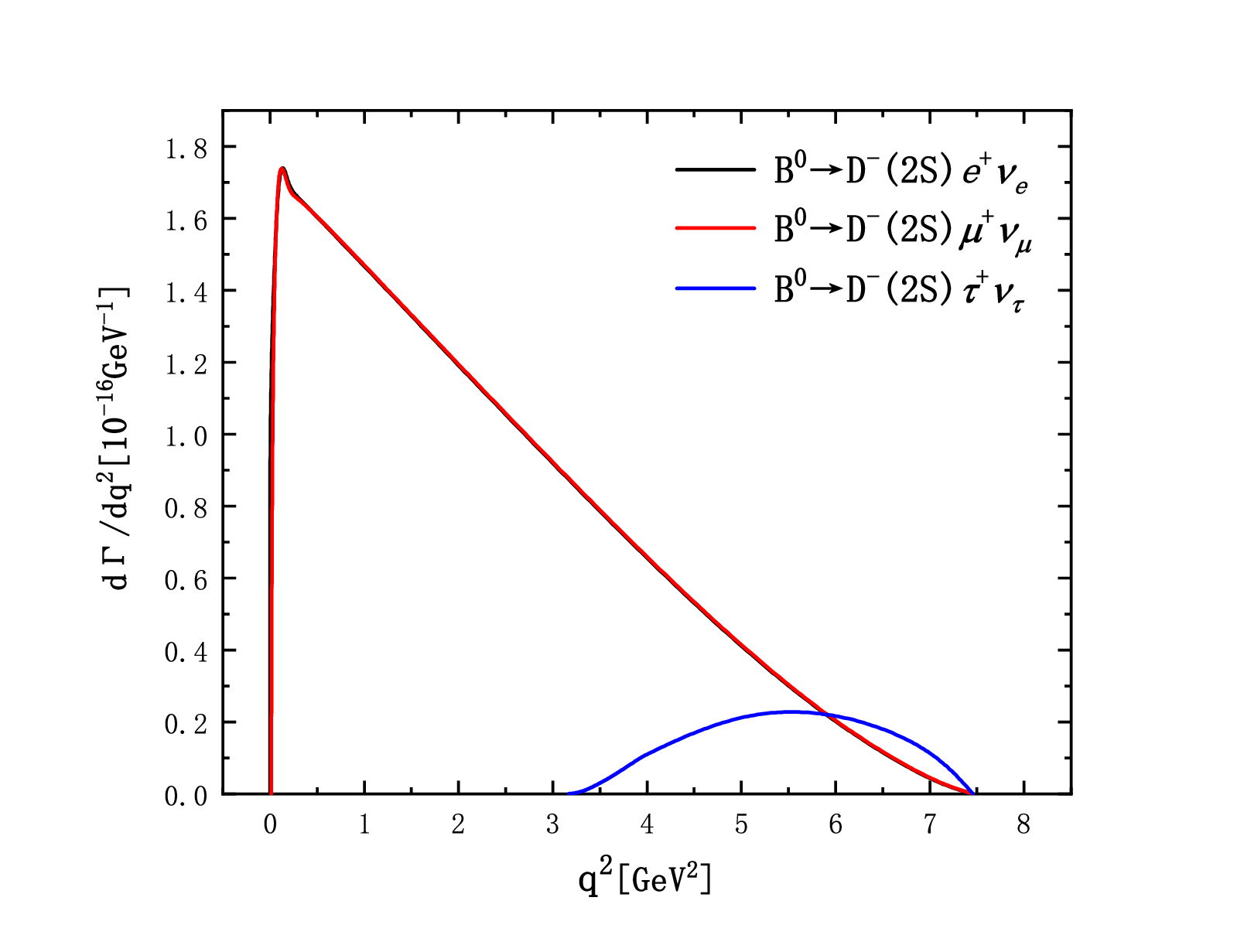}\quad}
	\subfigure[]{\includegraphics[width=0.4\textwidth]{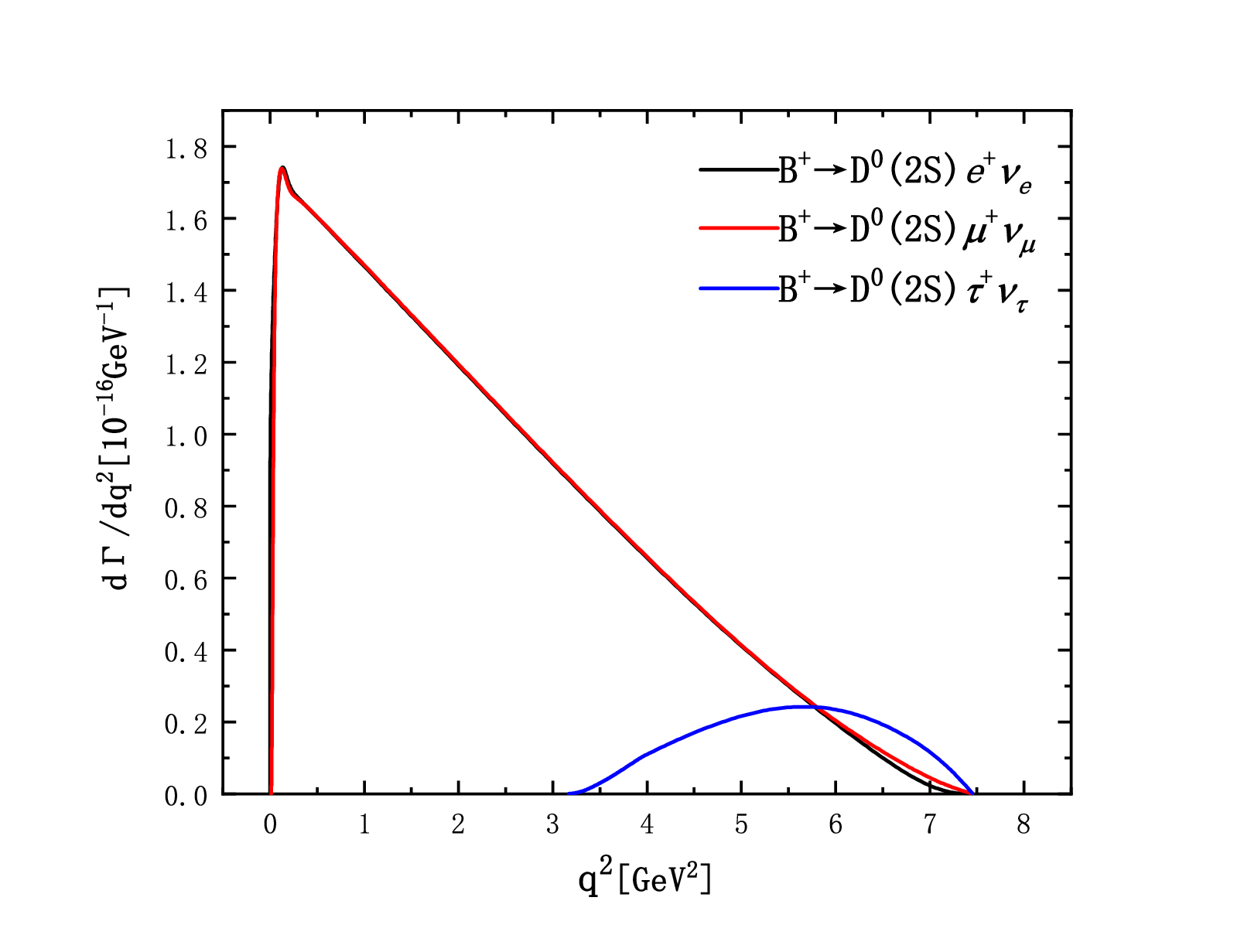}}\\
	\subfigure[]{\includegraphics[width=0.4\textwidth]{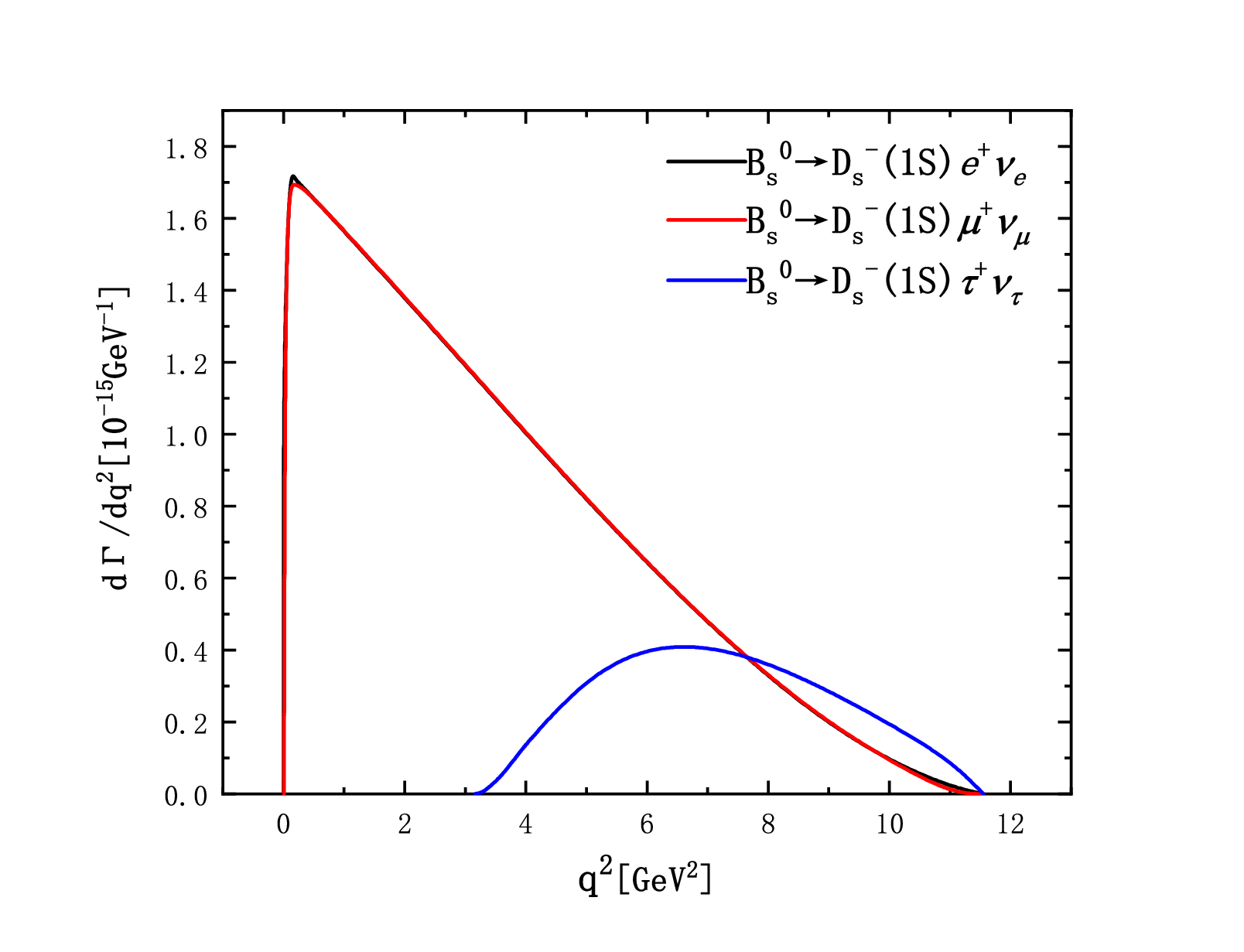}\quad}
	\subfigure[]{\includegraphics[width=0.4\textwidth]{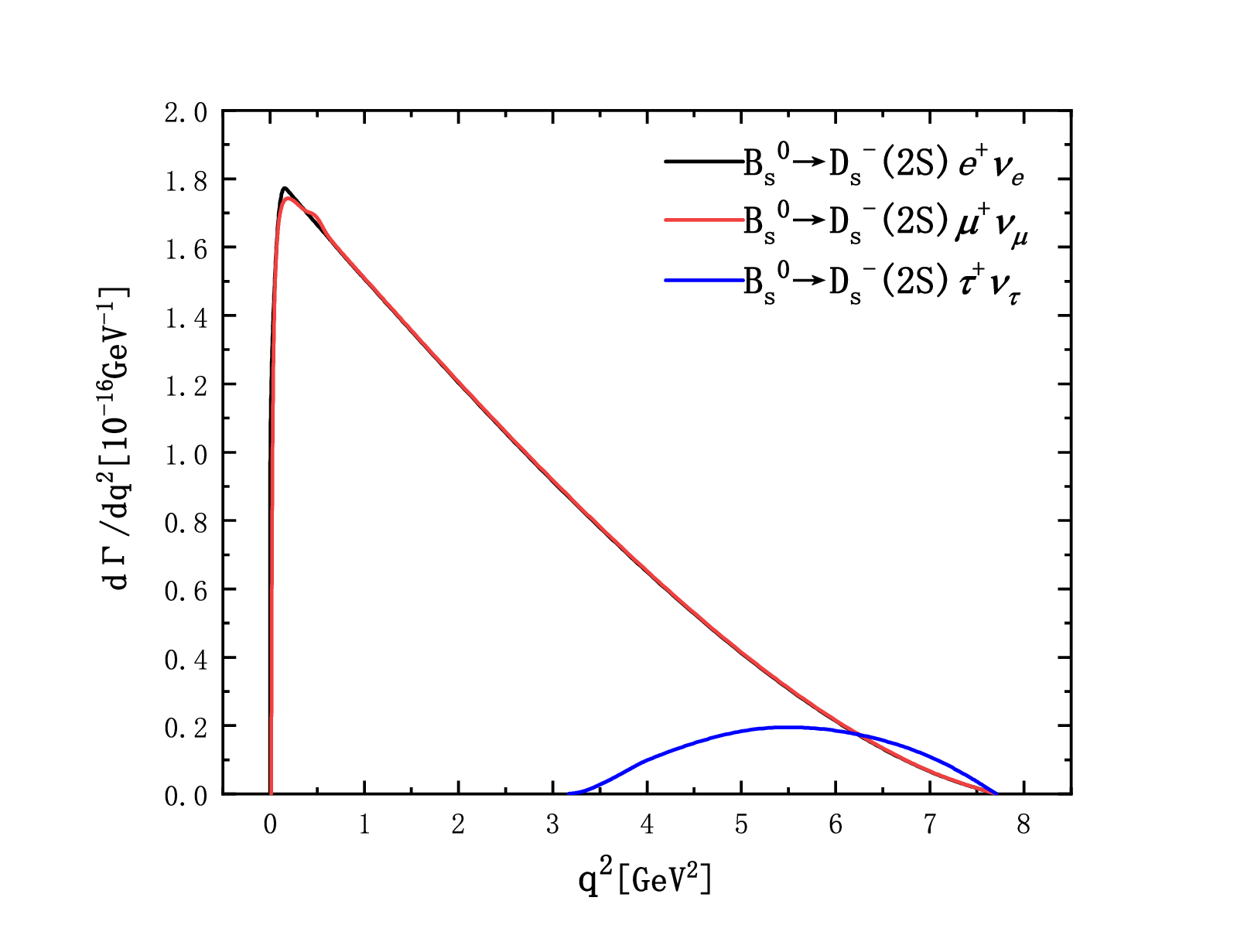}}\\
	\caption{The $q^{2}$ dependence of the differential decay rates $d\Gamma/dq^{2}$ for the decay $B^0 \to D^-(1S)\ell^{+}\nu_{\ell}$ (a), $B^+ \to D^0(1S)\ell^{+}\nu_{\ell}$ (b), $B^0 \to D^-(2S)\ell^{+}\nu_{\ell}$ (c), $B^+ \to D^0(2S)\ell^{+}\nu_{\ell}$ (d), $B^0_s \to D^-_s(1S)\ell^{+}\nu_{\ell}$ (e) and $B^0_s \to D^0_s(2S)\ell^{+}\nu_{\ell}$ (f).}\label{fig:T4}
\end{figure}

\begin{figure}[H]
	\vspace{0.4cm}
	\centering
	\subfigure[]{\includegraphics[width=0.4\textwidth]{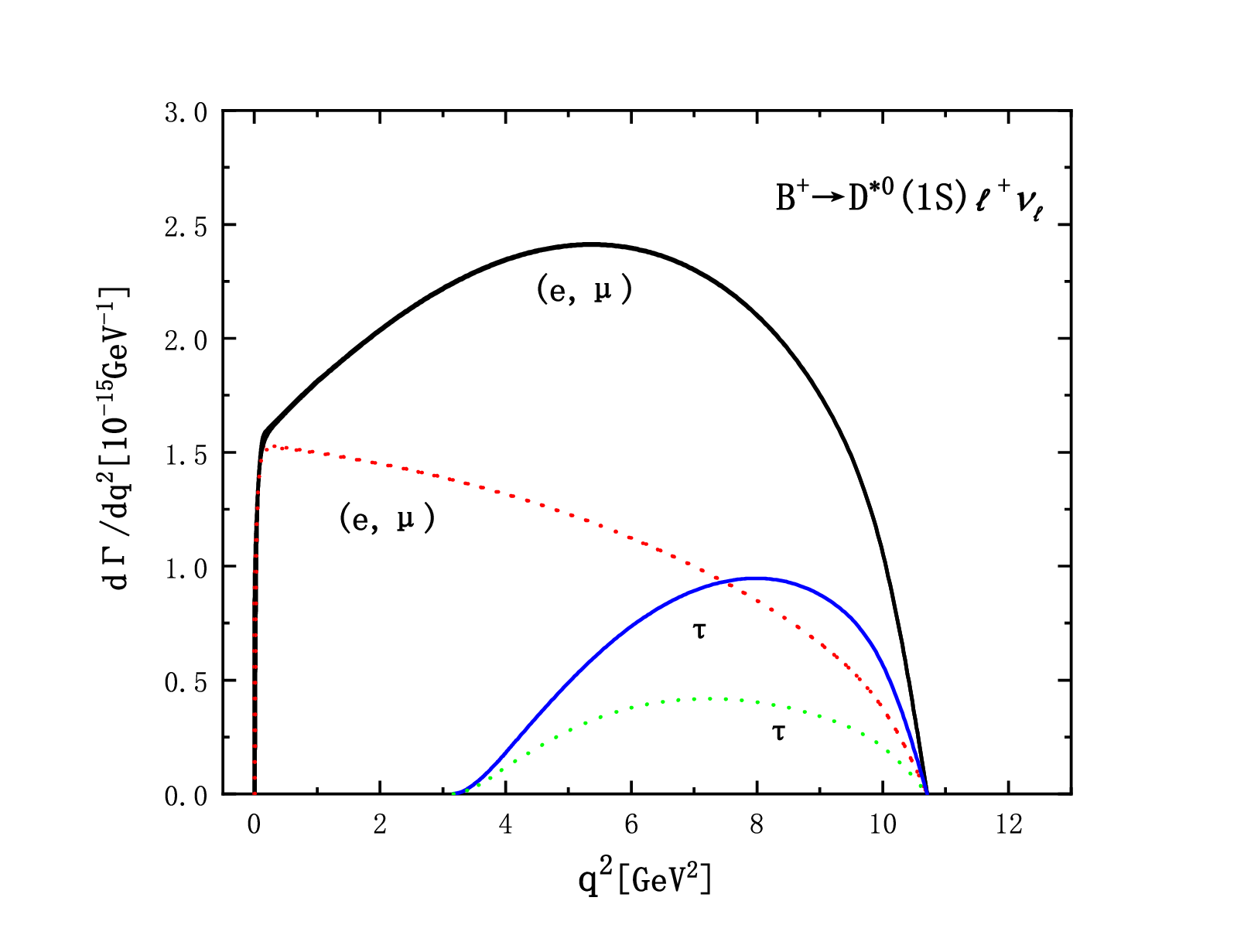}\quad}
	\subfigure[]{\includegraphics[width=0.4\textwidth]{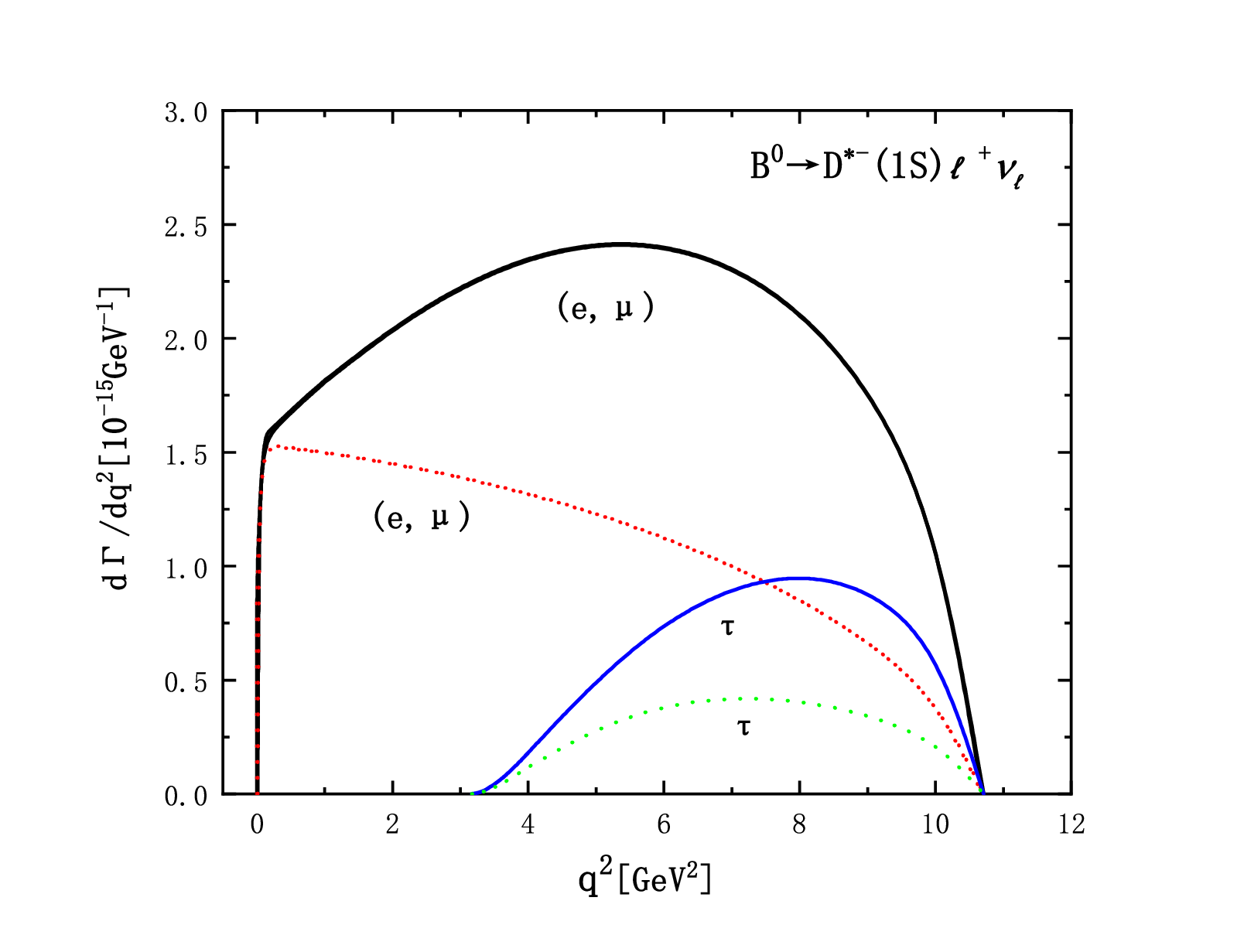}}\\
	\subfigure[]{\includegraphics[width=0.4\textwidth]{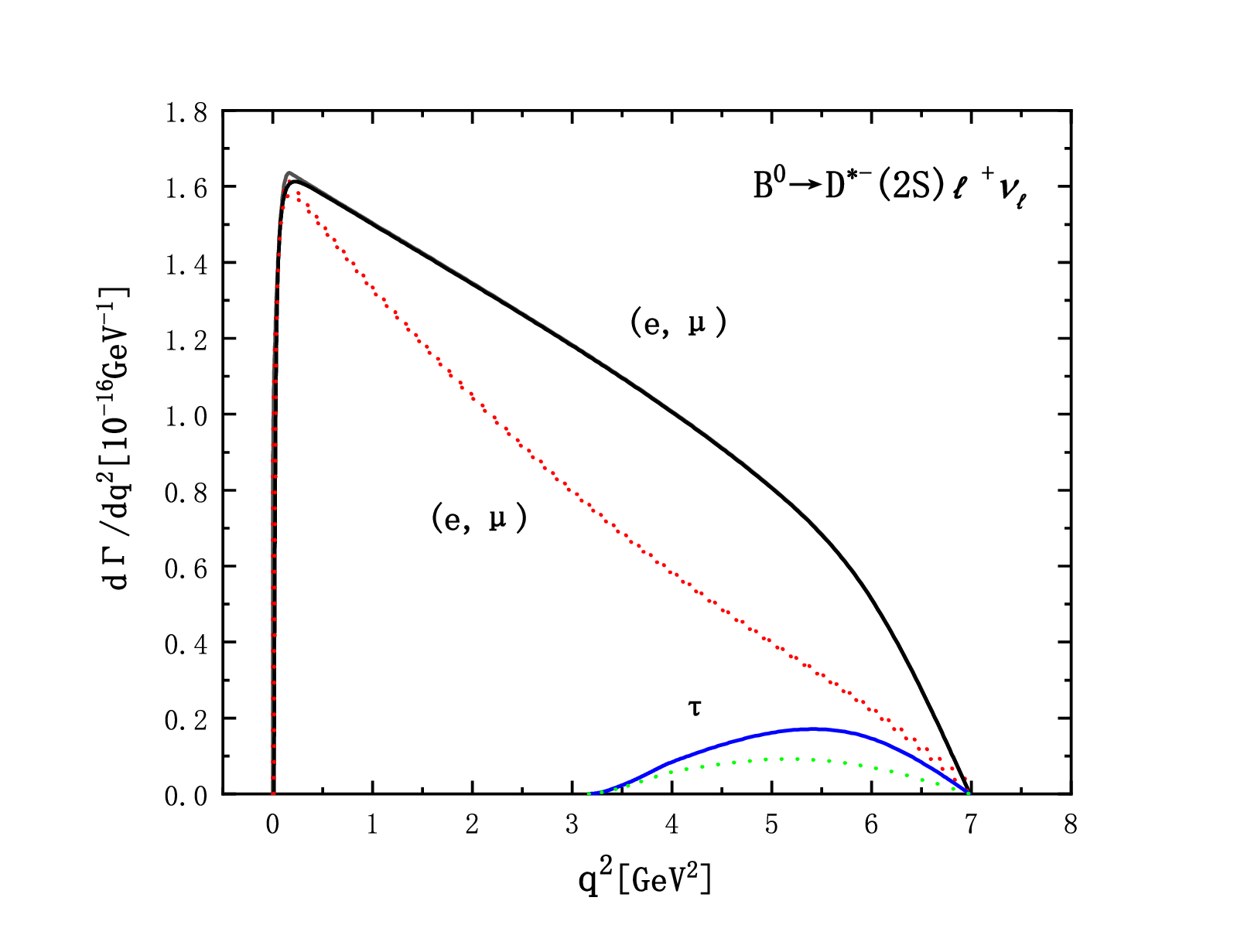}\quad}
	\subfigure[]{\includegraphics[width=0.4\textwidth]{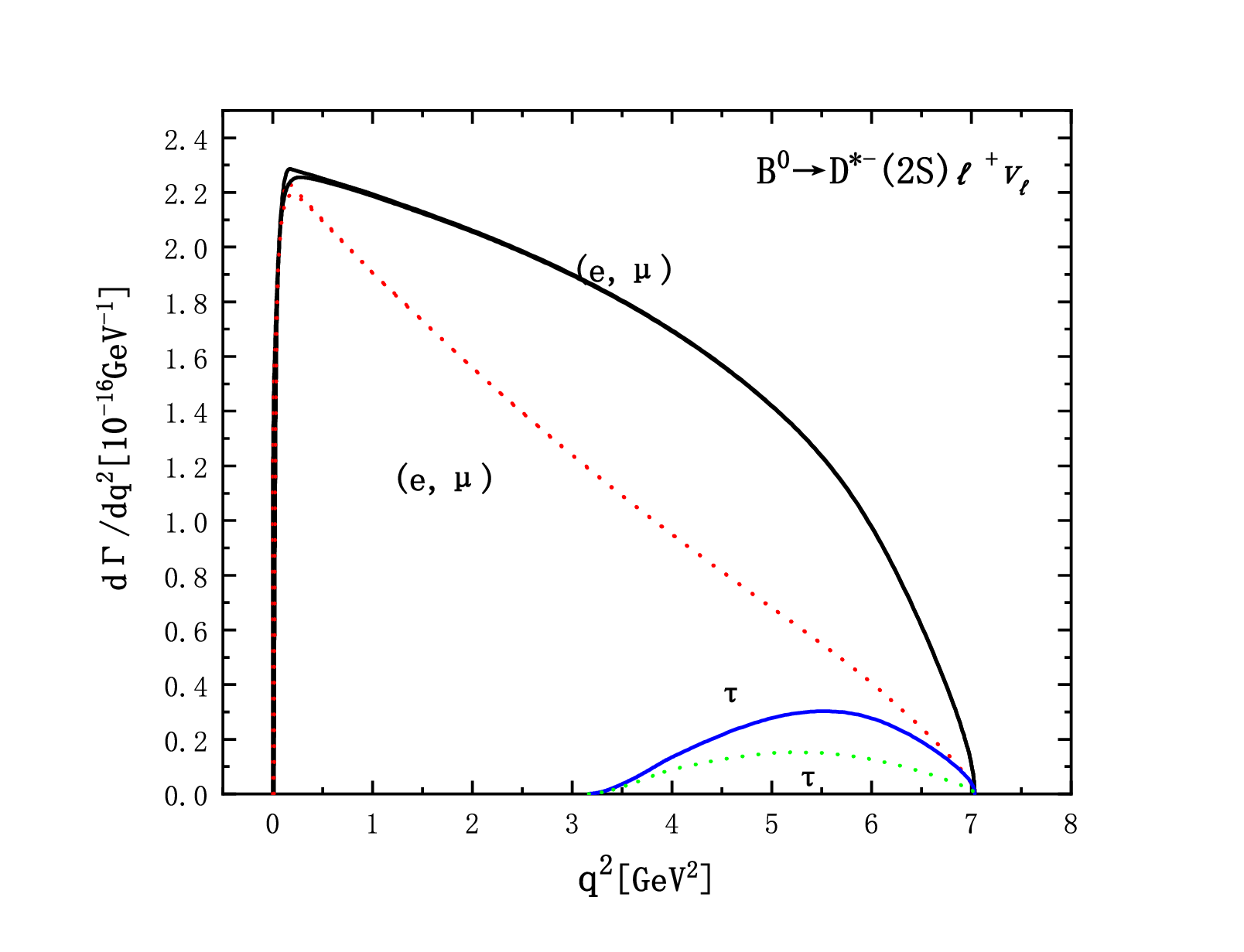}}\\
	\subfigure[]{\includegraphics[width=0.4\textwidth]{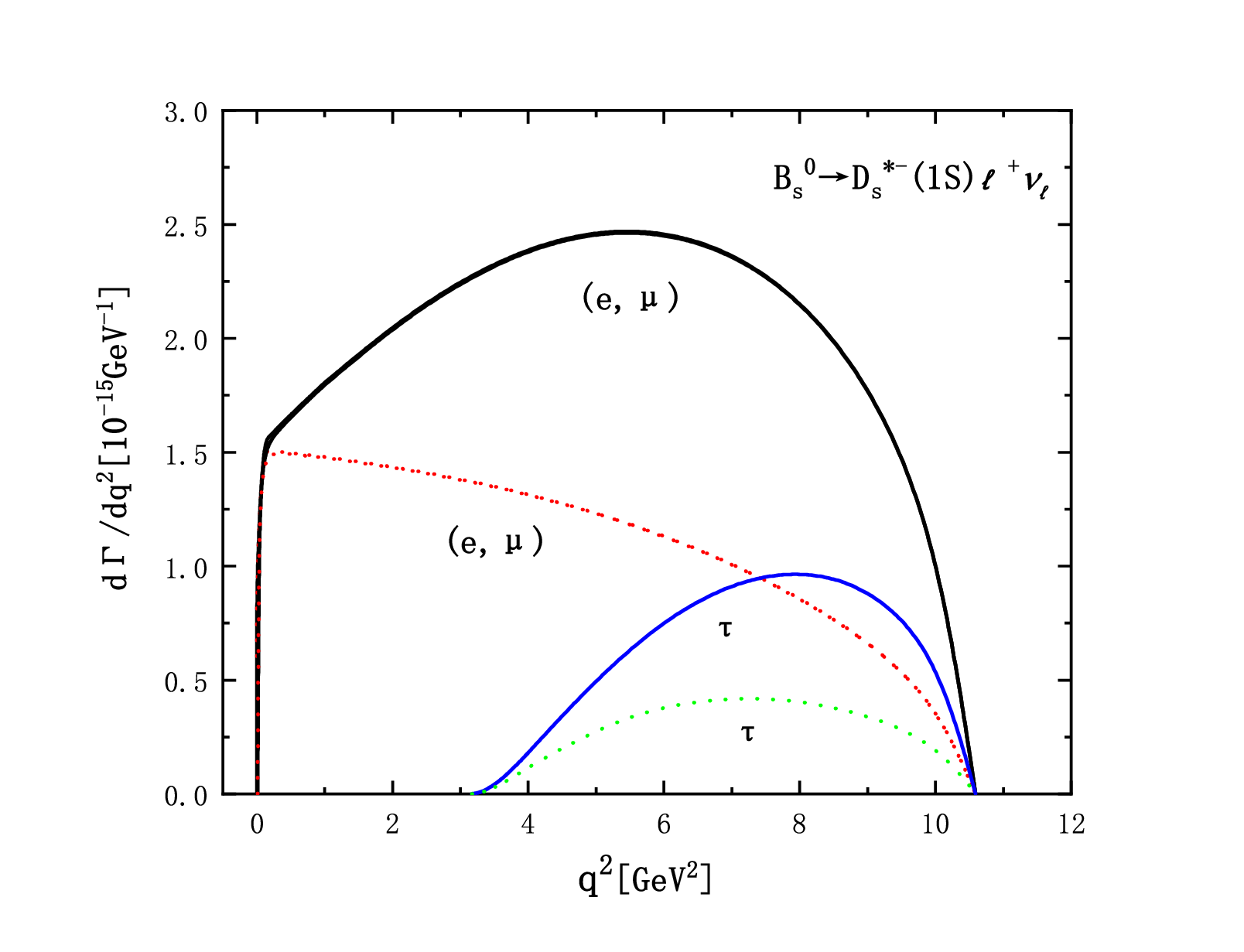}\quad}
	\subfigure[]{\includegraphics[width=0.4\textwidth]{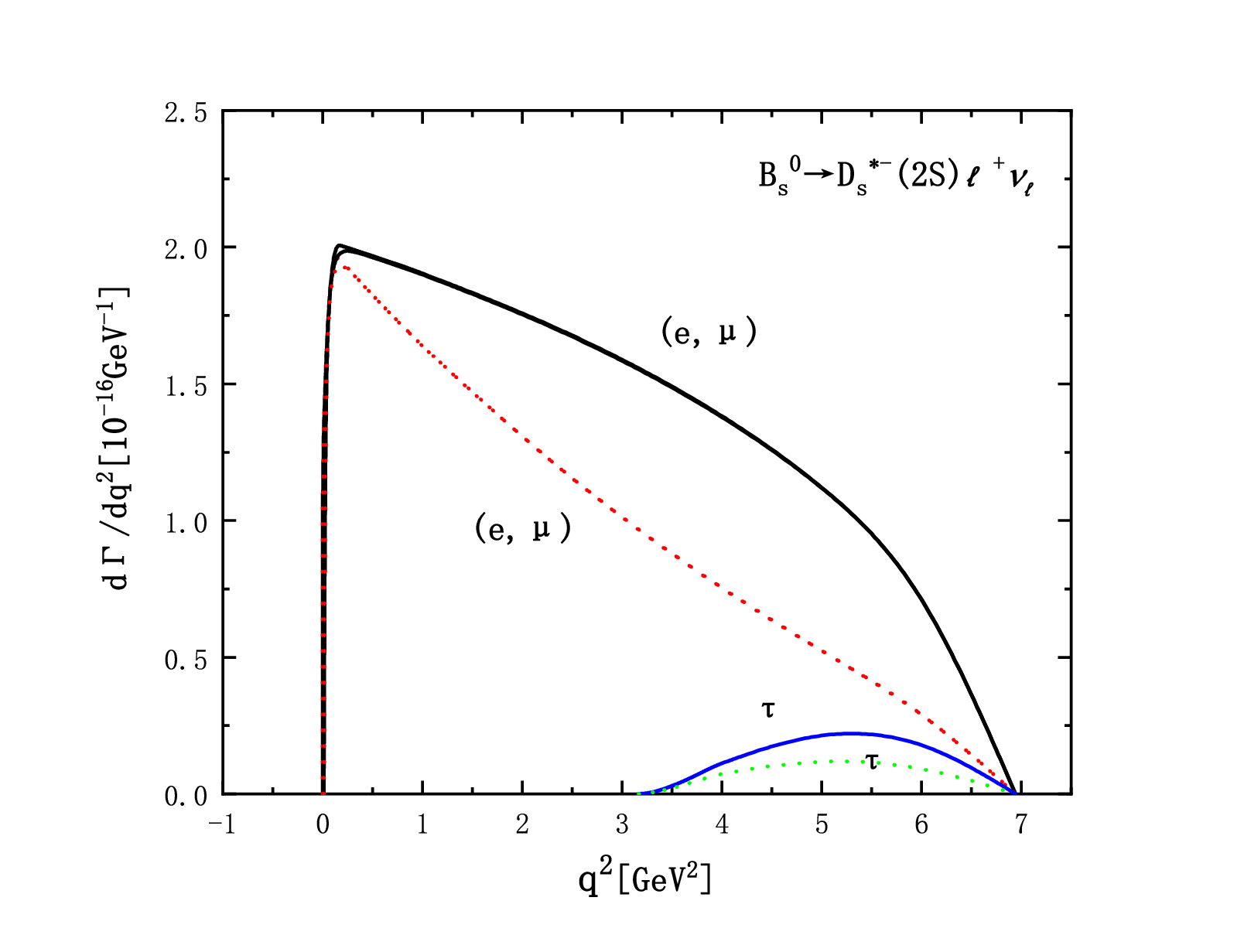}}\\
	\caption{The  $q^2$ dependence of the differential decay rates $d\Gamma/dq^2$ (the solid lines) and $d\Gamma^{L}/dq^2$ (the dashed lines refer to $\ell^{+\prime}\nu_{\ell^\prime}$ mode, the dotted lines represent for $\tau^+\nu_\tau$ mode) for the decays $B^+\to D^{*0}(1S)\ell^{+}\nu_{\ell}$ (a), $B^0\to D^{*-}(1S)\ell^{+}\nu_{\ell}$ (b), $B^+\to D^{*0}(2S)\ell^{+}\nu_{\ell}$ (c), $B^0\to D^{*-}(2S)\ell^{+}\nu_{\ell}$ (d), $B^0_s\to D^{*-}_s(1S)\ell^{+}\nu_{\ell}$ (e) and $B^0_s\to D^{*-}_s(2S)\ell^{+}\nu_{\ell}$ (f).}\label{fig:T5}
\end{figure}
%%%%%%%%%%%%%%%%%%%%%%%%%%%%%%%%%%%%%%%%%%%%%%%%%%%%%%%%%%%%%%%%%%%%%%%%
%                               references
%%%%%%%%%%%%%%%%%%%%%%%%%%%%%%%%%%%%%%%%%%%%%%%%%%%%%%%%%%%%%%%%%%%%%%%%

\end{document}